\documentclass[a4paper,fleqn,usenatbib]{mnras}

\usepackage[T1]{fontenc}
\usepackage{ae,aecompl}
\usepackage{graphicx}
\usepackage{amsmath}

\usepackage{amssymb}
\usepackage{newtxtext,newtxmath}
\usepackage{color}
\usepackage{soul}

\newcommand{\kms}{\hbox{km s$^{-1}$}}
\newcommand{\vsini}{\mbox{$v\,\sin\,i$}}
\newcommand{\zdi}{\textsc{zdi}}
\newcommand{\zdots}{\textsc{zdots}}

\title[Magnetic topologies of TAP~4 and TAP~40]{Magnetic topologies of two weak-line T Tauri stars TAP~4 and TAP~40}
\author[Y.~Xiang et al.]{Yue Xiang,$^{1,2}$\thanks{E-mails: xy@ynao.ac.cn (YX); shenghonggu@ynao.ac.cn(SG)} Shenghong Gu,$^{1,2,3\star}$ J.-F. Donati,$^{4}$ G. A. J. Hussain,$^{5}$ A. Collier Cameron$^{6}$ \and and the MaTYSSE collaboration\\
$^{1}$Yunnan Observatories, Chinese Academy of Sciences, Kunming 650216, China\\
$^{2}$Key Laboratory for the Structure and Evolution of Celestial Objects, Chinese Academy of Sciences, Kunming 650216, China\\
$^{3}$School of Astronomy and Space Science, University of Chinese Academy of Sciences, Beijing 101408, China\\
$^{4}$IRAP, Universit\'{e} de Toulouse, CNRS, UPS, CNES, F-31400 Toulouse, France\\
$^{5}$Science Division, European Space Research and Technology Centre (ESA/ESTEC), Keplerlaan 1, NL-2201AZ Noordwijk, the Netherlands\\
$^{6}$School of Physics and Astronomy, University of St Andrews, Fife KY16 9SS, UK}

\date{Accepted XXX. Received YYY; in original form ZZZ}
\pubyear{2022}

\begin{document}
\label{firstpage}
\pagerange{\pageref{firstpage}--\pageref{lastpage}}
\maketitle

\begin{abstract}
We present a Zeeman-Doppler imaging study of two weak-line T Tauri stars TAP~4 and TAP~40, based on the high-resolution spectropolarimetric observations with ESPaDOnS at the Canada-France-Hawaii Telescope in November 2013, in the framework of the MaTYSSE large programme. We apply two Zeeman-Doppler imaging codes to the Stokes I and V profiles to reconstruct their brightness and large-scale magnetic field images. The results given by the two imaging codes are in good agreement with each other. TAP~4 shows a large polar cool spot and several intermediate-latitude warm spots on its surface, whereas TAP~40 exhibits very weak variations in its Stokes I profiles suggesting a mostly unspotted photosphere. We detect Zeeman signatures in the Stokes V profiles of both stars. The reconstructed magnetic maps reveal dominantly toroidal fields, which enclose about 60 per cent of the total magnetic energy for both of TAP~4 and TAP~40. Both stars show prominent circular ring features of the azimuthal magnetic field. We derive a solar-like surface differential rotation on TAP~4 from the tomographic modelling. The brightness image of TAP~4 is used to predict the radial velocity jitters induced by its activity. After filtering out the activity jitter, the RMS of its RVs is reduced from 1.7 \kms\ to 0.2 \kms, but we do not detect any periodic signals in the filtered RVs of TAP~4, implying that it is unlikely to host a close-in exoplanet more massive than $\sim$3.5~M$_{\rm Jup}$ at 0.1~au.
\end{abstract}

\begin{keywords}
techniques: polarimetric --
stars: imaging --
stars: magnetic fields --
stars: individual: TAP~4 --
stars: individual: TAP~40
\end{keywords}

\section{Introduction}

Magnetic fields are believed to play a significant role in the formation and evolution of stars and their surrounding planets. Pre-main sequence (PMS) cool stars show high levels of magnetic activity due to their high rotation rates. Such low-mass stars preserve a high angular momentum content and they will gradually dissipate it due to magnetic braking. The magnetic field also affects the accretion process of the PMS stars (see \citealt{donati2009} and the references therein).

Zeeman-Doppler imaging, which was first proposed by \citet{semel1989}, takes advantage of a series of time-resolved polarized spectra to recover the magnetic field structure on the stellar surface. It is a powerful tool for the investigation on the properties of magnetic field of young fast rotating stars at different evolutionary phase, which is important for understanding the evolution of the stellar magnetic field. Magnetic field topologies of Sun-like stars at different evolutionary stages are revealed through Zeeman-Doppler imaging (e.g. \citealt{hackman2016,rosen2016}), but the numbers of systems studied have remained limited up to now.

The Magnetic Topologies of Young Stars and Survival of close-in giant Exoplanets (MaTYSSE) large programme aims to monitor the magnetic field of the young T Tauri stars and search for the potential close-in hot Jupiters around them, using the high-resolution spectropolarimetric observations with multiple telescopes \citep{donati2014} and Zeeman-Doppler imaging. Through this project, the large-scale magnetic field topologies of a set of young active stars have been recovered (e.g. \citealt{donati2015,hill2017,donati2019}), and the hot Jupiters (hJs) surrounding two very young stars, V830 Tau \citep{donati2016} and TAP 26 \citep{yu2017}, have been detected to date, which have significant impact on the study of planet formation.

The weak-line T Tauri stars (wTTSs) are young low-mass stars whose inner discs are already dispersed, which is one of the main differences with respect to their progenitors, the classical T Tauri stars (cTTSs). TAP~4 and TAP~40 are two active young wTTSs in Taurus-Auriga \citep{basri1991,grankin2008}. Previous analysis of photometric data indicates that TAP~4 has a short rotational period of 0.482 d and TAP~40 rotates more slowly with a rotational period of 1.555~d \citep{grankin2008,grankin2013}.

In this work, we present tomographic modelling of TAP~4 and TAP~40, using two different Zeeman-Doppler imaging codes. We describe the spectropolarimetric observations and data reduction in Section 2, and discuss the evolutionary status and parameters of two stars in Section 3. The tomographic modelling for two stars are conducted in Section 4, including the brightness and magnetic field reconstructions, the surface differential rotation determination and the radial velocity (RV) filtering. We finally discuss and summarise the results in Section 5.

\section{Observations and data reduction}

The high-resolution spectropolarimetric observations of TAP~4 and TAP~40 were carried out in November 2013 using the spectropolarimeter ESPaDOnS mounted on the 3.6-m Canada-France-Hawaii Telescope (CFHT). ESPaDOnS has a spectral resolving power of 65~000 and a wavelength coverage from 370 to 1~000 nm in spectropolarimeter mode. A total of 31 Stokes I and Stokes V spectra of TAP~4 and 18 spectra of TAP~40 were collected from November 15 to 26, which covered about 23 and 7 rotational cycles of TAP~4 and TAP~40, respectively. We typically obtained 2--4 spectra per night for TAP~4 and 1--2 spectra per night for TAP~40. We summarize the observations, including the UT date, barycentric Julian Date (BJD), peak signal-to-noise ratio (SNR) and rotational cycle/phase in Table \ref{tab:log}. The raw data were reduced with LIBRE ESPRIT, the dedicated ESPaDOnS pipeline that produces 1D Stokes I and V spectra from the raw ESPaDOnS images. The rotational cycles of TAP~4 and TAP~40 were respectively calculated from the following ephemerides:
\begin{align}
\label{eq:eph}
\begin{split}
\textrm{BJD(d)} = 2456611.83 + 0.482E
\\
\textrm{BJD(d)} = 2456611.96 + 1.585E
\end{split}
\end{align}
where the start points were arbitrarily chosen. The rotational periods were taken from \citet{grankin2008}, but the value of TAP~40 has been fine-tuned for a better fit (see Section~4).

\begin{table}
\center
 \caption{Summary of observations.}
 \label{tab:log}
 \begin{tabular}{lccccc}
  \hline
  UT Date   & BJD$_{\rm TT}$ & SNR & SNR    & SNR  & Cycle/Phase\\
            & 2450000+       & raw &  LSD-I & LSD-V        &      \\
  \hline 
TAP~4 &&&&&\\
2013-11-15 & 6611.83559 & 103 &  888 & 2007 &  0.012\\
2013-11-15 & 6611.93441 &  89 &  860 & 1759 &  0.217\\
2013-11-15 & 6612.04650 &  84 &  819 & 1772 &  0.449\\
2013-11-16 & 6612.76210 & 101 &  878 & 1918 &  1.934\\
2013-11-16 & 6613.01104 &  97 &  860 & 1825 &  2.450\\
2013-11-17 & 6613.75459 &  97 &  894 & 1835 &  3.993\\
2013-11-17 & 6613.86670 & 108 &  947 & 2169 &  4.226\\
2013-11-17 & 6613.99974 & 100 &  904 & 1917 &  4.502\\
2013-11-17 & 6614.08710 &  96 &  887 & 1823 &  4.683\\
2013-11-18 & 6614.81879 &  93 &  874 & 1805 &  6.201\\
2013-11-19 & 6615.79046 & 101 &  899 & 1941 &  8.217\\
2013-11-19 & 6615.91156 &  98 &  888 & 1880 &  8.468\\
2013-11-19 & 6615.97844 &  97 &  867 & 1855 &  8.607\\
2013-11-19 & 6616.04393 &  97 &  875 & 1861 &  8.743\\
2013-11-20 & 6616.73490 &  84 &  810 & 1848 & 10.176\\
2013-11-20 & 6616.84261 &  96 &  858 & 1777 & 10.400\\
2013-11-20 & 6616.95424 &  90 &  854 & 1756 & 10.631\\
2013-11-20 & 6617.10569 &  78 &  607 & 2304 & 10.945\\
2013-11-23 & 6619.75070 &  91 &  866 & 1848 & 16.433\\
2013-11-23 & 6619.86170 &  88 &  863 & 1799 & 16.663\\
2013-11-23 & 6619.97877 & 114 &  967 & 2323 & 16.906\\
2013-11-23 & 6620.04474 & 111 &  959 & 2215 & 17.043\\
2013-11-24 & 6620.87867 & 102 &  909 & 1939 & 18.773\\
2013-11-25 & 6621.78141 & 107 &  904 & 2084 & 20.646\\
2013-11-25 & 6621.88841 & 113 &  933 & 2289 & 20.868\\
2013-11-25 & 6621.95597 & 110 &  934 & 2212 & 21.008\\
2013-11-25 & 6622.06311 & 107 &  913 & 2088 & 21.231\\
2013-11-26 & 6622.76619 & 106 &  896 & 2017 & 22.689\\
2013-11-26 & 6622.83456 & 110 &  926 & 2198 & 22.831\\
2013-11-26 & 6622.89830 & 113 &  936 & 2287 & 22.963\\
2013-11-26 & 6623.01087 &  99 &  892 & 1903 & 23.197\\
&&&&&\\
TAP~40 &&&&&\\
2013-11-15 & 6611.96816 & 112 &  500 & 2062 &  0.005\\
2013-11-15 & 6612.09788 & 137 &  499 & 2543 &  0.087\\
2013-11-16 & 6612.92022 & 144 &  528 & 2759 &  0.606\\
2013-11-16 & 6613.08941 & 124 &  527 & 2295 &  0.713\\
2013-11-17 & 6613.83316 & 144 &  534 & 2766 &  1.182\\
2013-11-18 & 6614.89824 & 136 &  661 & 2712 &  1.854\\
2013-11-19 & 6615.86713 & 146 &  540 & 2813 &  2.465\\
2013-11-19 & 6616.01089 & 131 &  539 & 2433 &  2.556\\
2013-11-20 & 6616.91928 & 123 &  521 & 2278 &  3.129\\
2013-11-20 & 6617.03109 &  78 &  522 & 1633 &  3.199\\
2013-11-20 & 6617.07290 &  94 &  544 & 1619 &  3.226\\
2013-11-23 & 6619.82874 & 128 &  524 & 2397 &  4.965\\
2013-11-23 & 6620.01186 & 149 &  528 & 2870 &  5.080\\
2013-11-24 & 6620.91183 & 133 &  524 & 2502 &  5.648\\
2013-11-25 & 6621.92131 & 148 &  527 & 2850 &  6.285\\
2013-11-25 & 6622.09622 & 143 &  527 & 2693 &  6.395\\
2013-11-26 & 6622.80209 & 141 &  524 & 2625 &  6.840\\
2013-11-26 & 6622.93164 & 147 &  527 & 2800 &  6.922\\

  \hline 
 \end{tabular}
\end{table}

It is well known that (Zeeman) Doppler imaging needs to use well-sampled time series of profiles with high SNRs. Thus, Least-Squares Deconvolution (LSD; \citealt{donati1997}), which extracts a high-SNR average profile from thousands of photoshperic lines, was applied to the observed Stokes I and V spectra. The stellar line lists used in the LSD process were obtained from a Kurucz model atmosphere \citep{kurucz1993} with $T_{\rm eff}$ = 5250 K and log $g$ = 4.5 for TAP~4 and the one with $T_{\rm eff}$ = 4750 K and $\log g$ = 4.5 for TAP~40. Lines in regions including telluric lines and strong chromospheric lines were removed from the lists. SNRs of the Stokes I and V LSD profiles are also listed in Table \ref{tab:log}. SNRs of Stokes I profiles were derived from their continuum windows. Different to that of Stokes V profiles, the noise of Stokes I profiles is dominated by residual line blending generated by LSD (e.g. inaccurate line list), rather than the photon noise, and thus the performance of LSD for Stokes I spectra is smaller than that for Stokes V spectra \citep{donati1997}. 

Several stellar spectra of TAP~4 suffered from the moon light contamination. We thus performed the two-step process, proposed by \citet{donati2016}, to remove lunar contamination from the polluted spectra, with the help of the Doppler imaging. Firstly, we performed Doppler imaging using the original Stokes I profiles excluding the polluted region, then fitted the residuals with Gaussian profiles. Secondly, we subtracted the fitted Gaussian profiles from the affected profiles and then derived the final Doppler images.

\begin{figure*}
\center
\includegraphics[width = .45\textwidth]{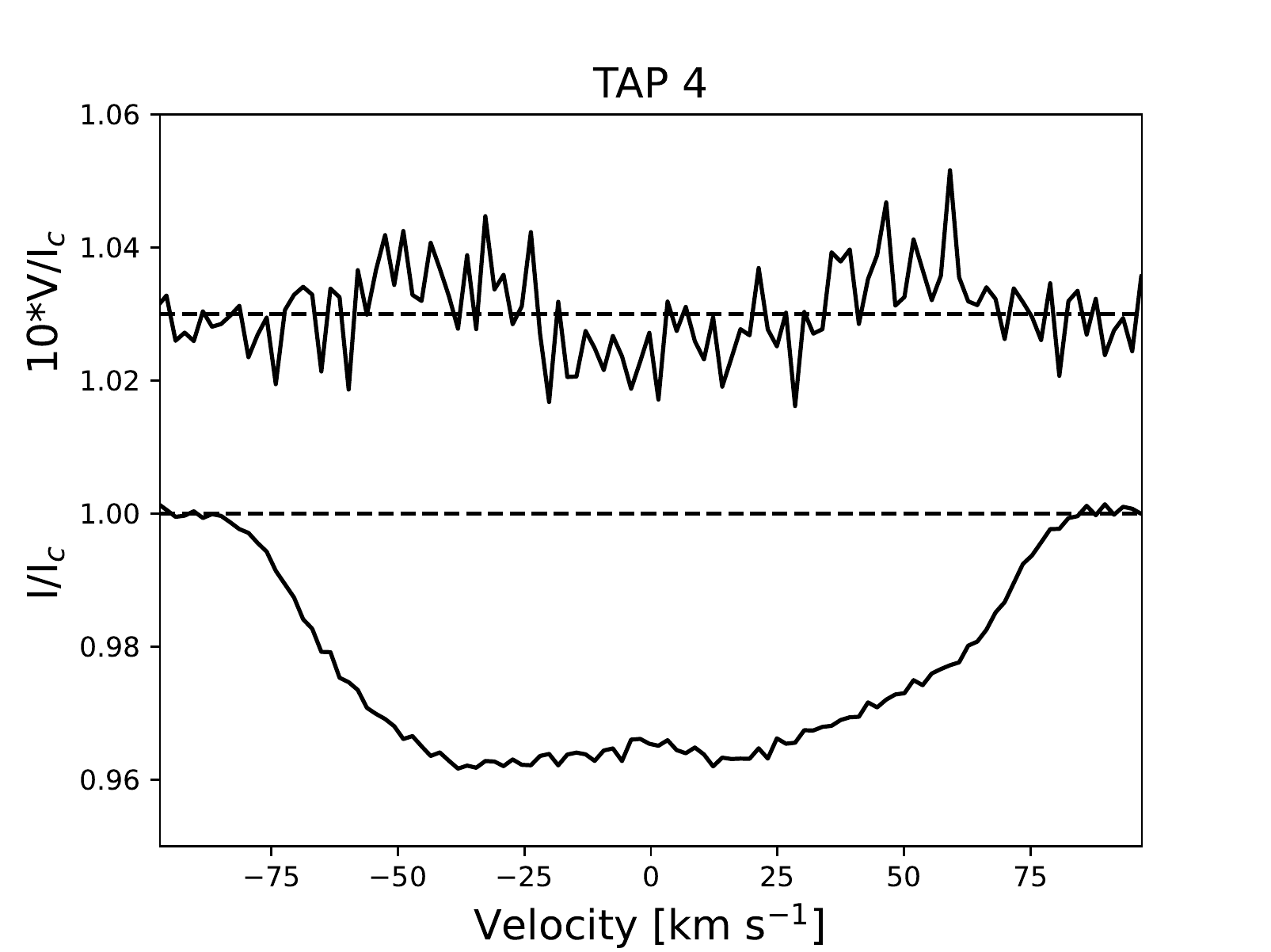}
\includegraphics[width = .45\textwidth]{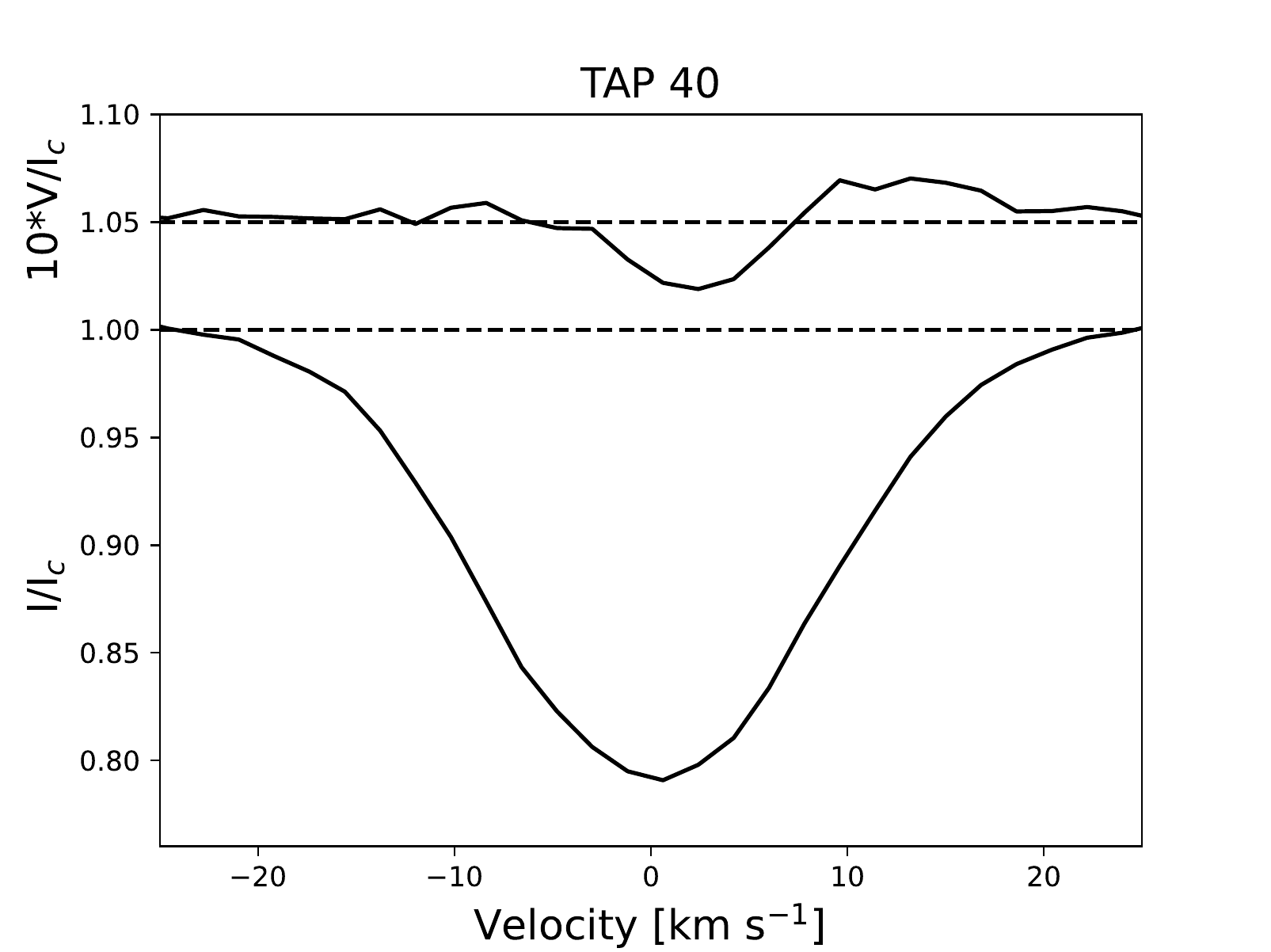}
\caption{Examples of the Stokes I and V profiles of TAP~4 and TAP~40.}
\label{fig:example}
\end{figure*}

\begin{figure*}
\center
\includegraphics[width = .31\textwidth]{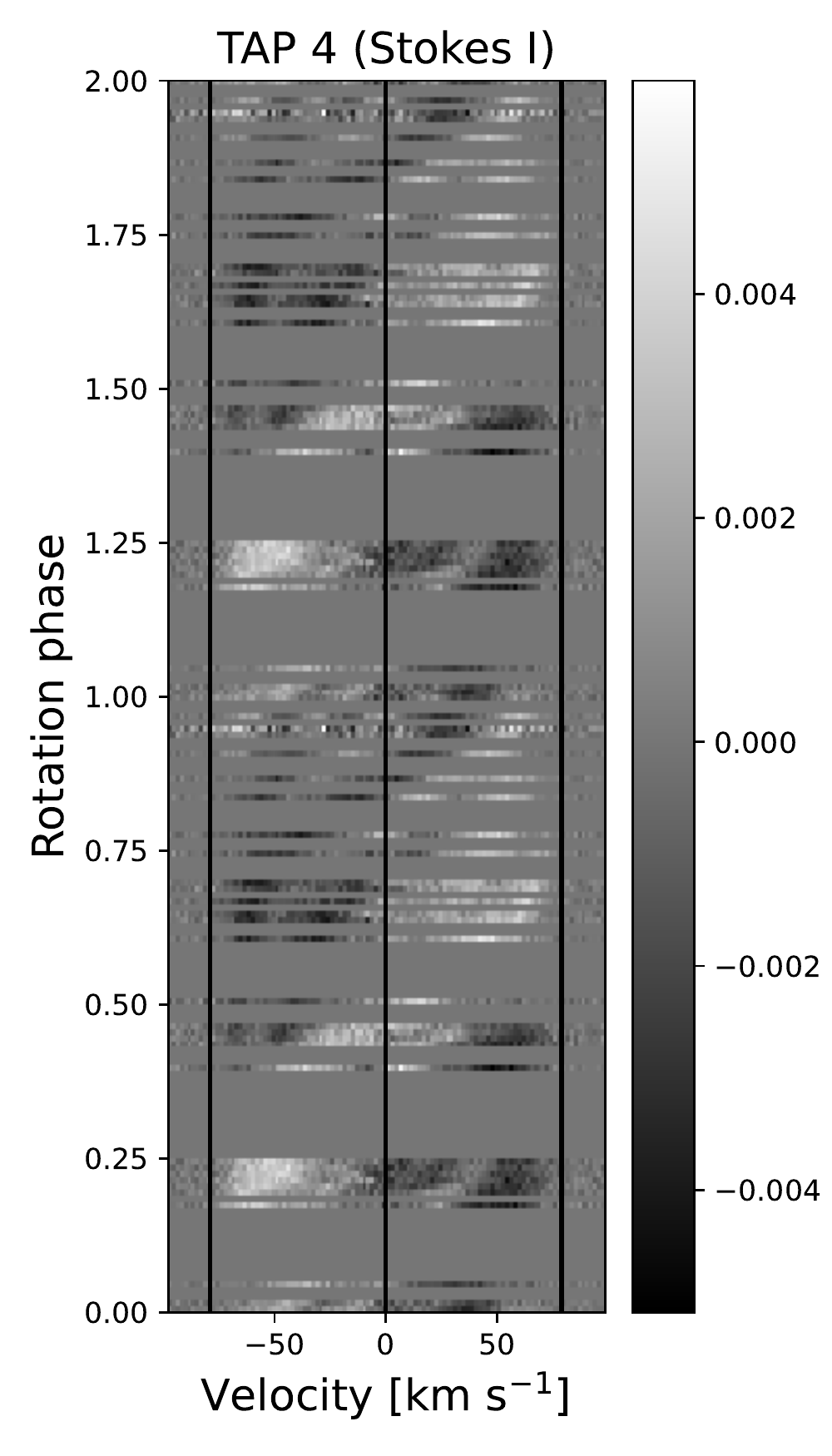}
\includegraphics[width = .31\textwidth]{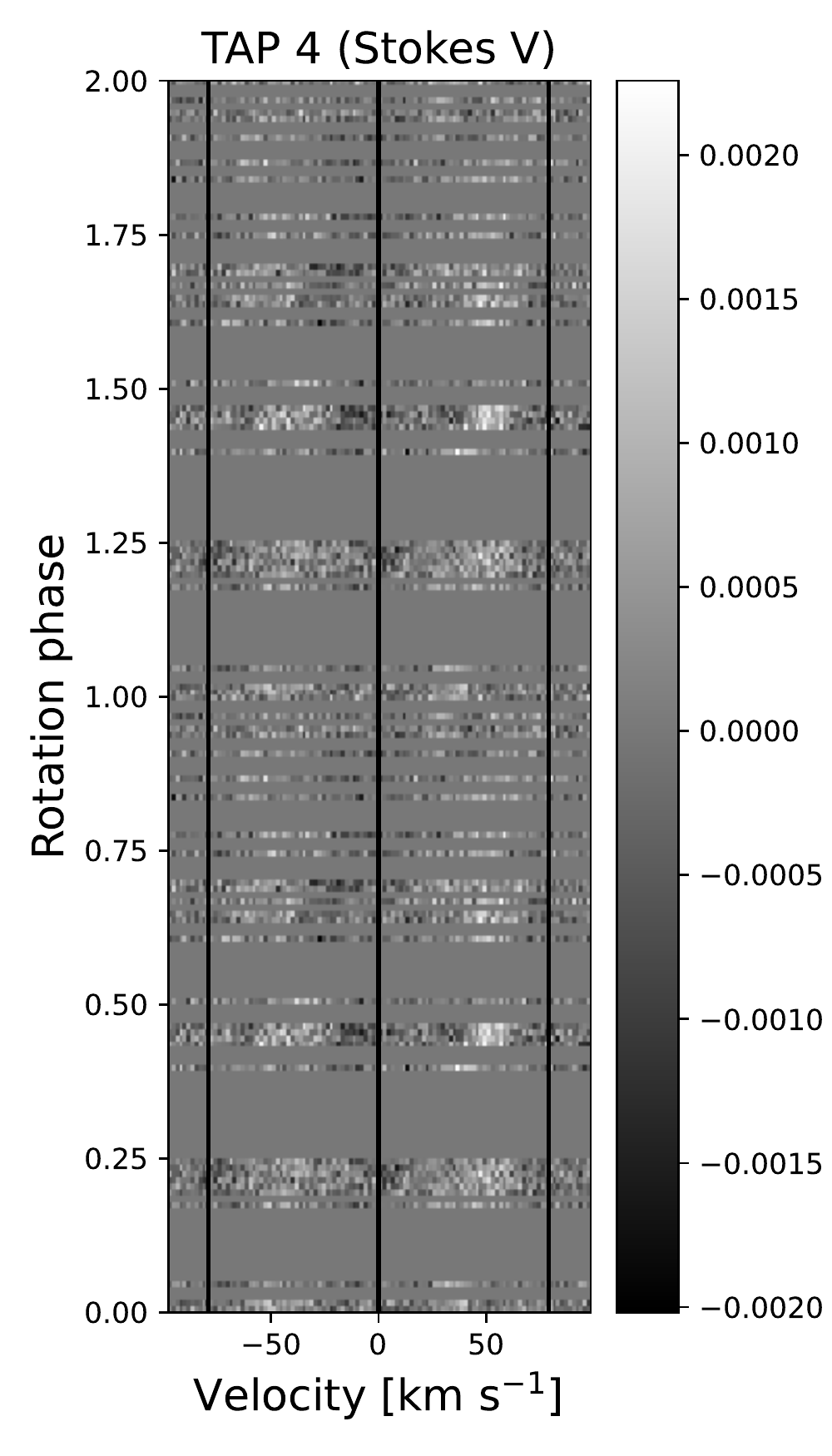}
\includegraphics[width = .31\textwidth]{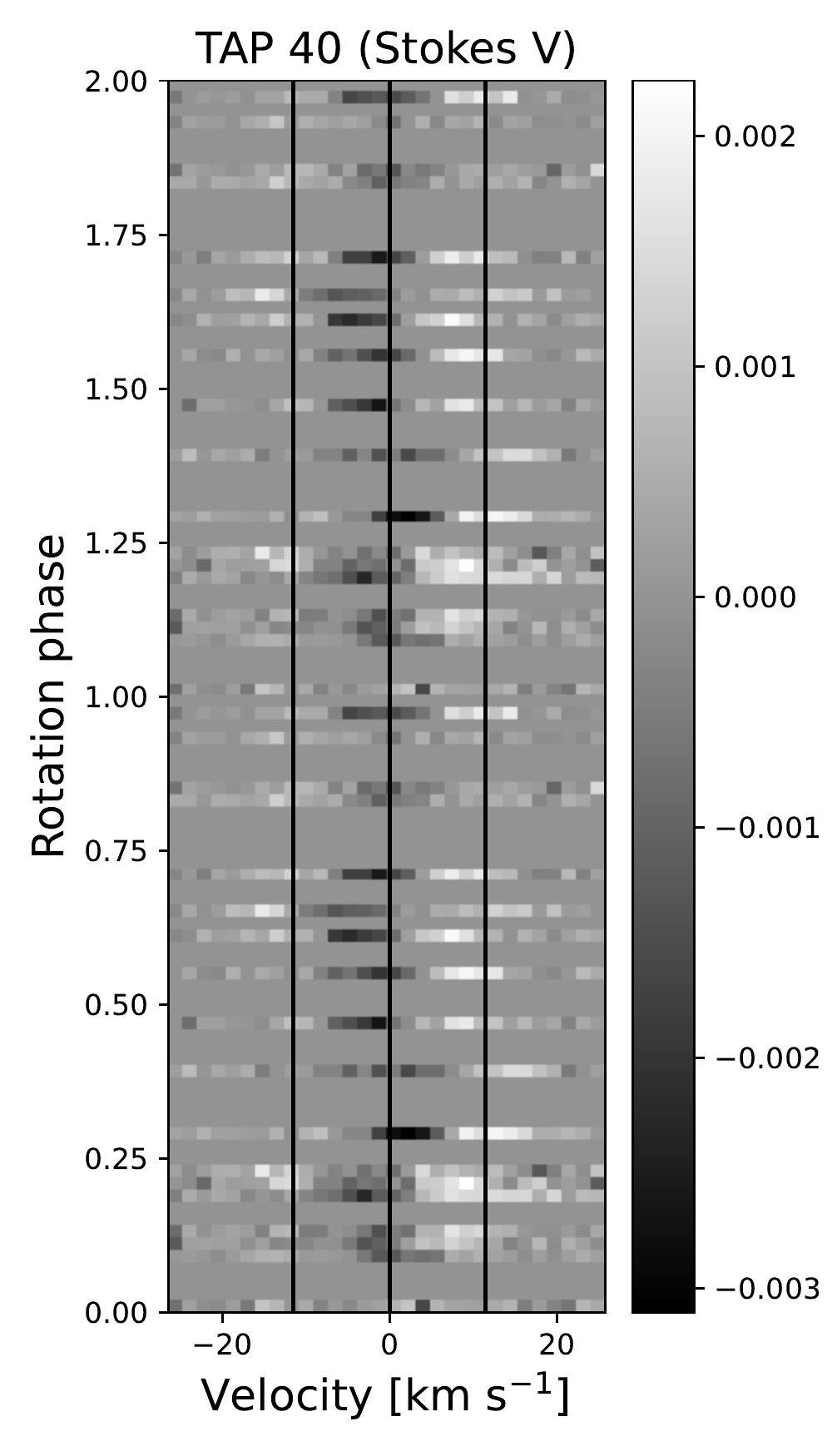}
\caption{Dynamic spectra of Stokes I (left) and V (right) profiles for TAP~4 and Stokes V profiles of TAP~40, phased using the rotation period in Equation \ref{eq:eph}. The average Stokes I profile was subtracted from Stokes I profile to emphasize spot signatures more clearly. The vertical lines represent $\pm v \sin i$ as well as the line centre in all panels.}
\label{fig:int}
\end{figure*}

Fig. \ref{fig:example} displays examples of LSD Stokes I and V profiles for TAP~4 and TAP~40. The Stokes I profile of the rapid rotator TAP~4 shows strong distortions and a flat bottom likely caused by a polar spot. The slower rotator TAP~40 shows clear Zeeman signatures in the Stokes V profile, but its unpolarized profile is nearly symmetric with respect to the line center. Fig. \ref{fig:int} shows a dynamic spectrum of the phased Stokes I profiles of TAP~4, where each profile is subtracted from the averaged profile of all observations. The dynamic Stokes I profiles of TAP~4 show clear spot signatures moving from blue to red wavelengths. The dynamic Stokes V profiles (see Fig. \ref{fig:int}) of TAP~4 and TAP~40 also show similar symmetric features, likely related to toroidal fields (e.g. \citealt{lehmann2022}).

\section{Evolutionary status and stellar parameters}

\begin{figure}
\center
\includegraphics[width = .45\textwidth]{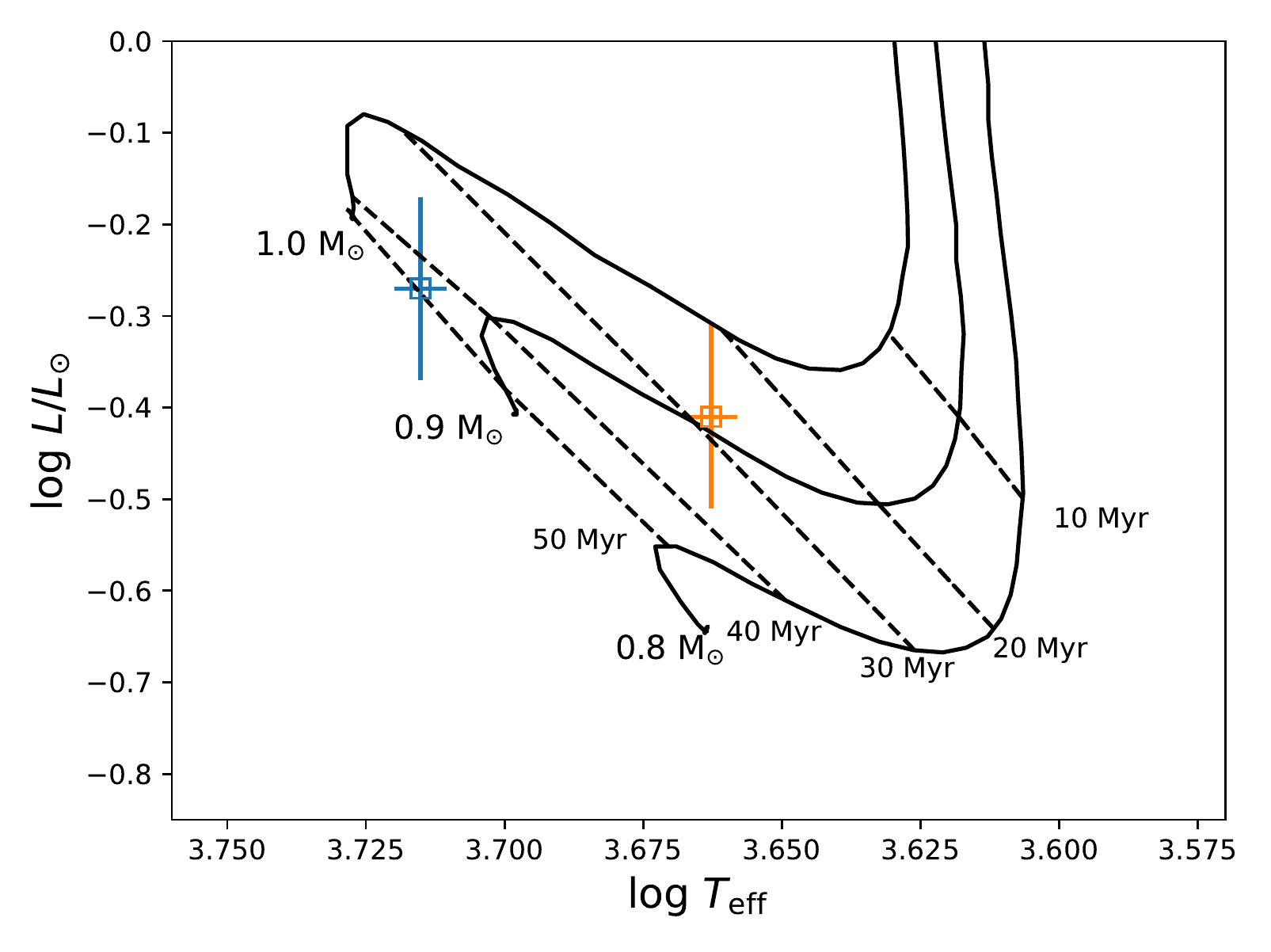}
\caption{Estimated locations of TAP~4 (blue open square) and TAP~40 (orange open square) in the HR diagram. The PMS evolutionary tracks are taken from \citet{siess2000}, assuming solar metallicity with convective overshooting.}
\label{fig:evl}
\end{figure}

The Gaia astrometric solution \citep{gaia2018,gaia2021,gaia2022} gives a parallax value of TAP~4, which is 6.878 $\pm$ 0.014~mas, corresponding to a distance of 145.4 $\pm$ 0.3 pc or a distance modulus of 5.813 $\pm$ 0.004. From long-term photometry, \citet{grankin2008} observed a maximum V magnitude of 12.0 and thus the V magnitude of the unspotted star may be 11.8 $\pm$ 0.2 assuming a spot coverage of $\sim$ 20 per cent on the visible stellar surface. The effective temperature and logarithmic gravity of TAP~4 were derived using the spectral classification tool developed by \citet{donati2012}. The results are $T_{\rm eff} = 5190 \pm 50$ K and $\log g = 4.6 \pm 0.2$ in cgs units. The B-V index expected at this temperature is equal to 0.79~$\pm$~0.02 \citep{pecaut2013} and the observed averaged value is equal to 0.88~$\pm$~0.05 \citep{grankin2008}, thus we derived that the visual extinction A$_{V}$ is equal to 0.28 $\pm$ 0.15. The bolometric correction BC$_{V}$ is equal to -0.27 $\pm$ 0.05 at this temperature \citep{pecaut2013}. Hence the absolute bolometric magnitude M$_{\textrm{bol}}$ is equal to 5.44~$\pm$~0.27 and thus the luminosity of TAP~4 is $\log (L/L_{\odot}$) = $-0.27~\pm~0.11$. Then we derived a stellar mass of 0.95~$\pm$~0.05~M$_{\odot}$, a radius of 0.9~$\pm$~0.1 R$_{\odot}$, and a stellar age of about 47~Myr for TAP~4 from comparing the position of TAP~4 in the H-R diagram with the PMS evolutionary tracks of \citet{siess2000} assuming solar abundance with convective overshooting (see Fig. \ref{fig:evl}). Given the rotational period and $v \sin i$ of TAP~4, we can infer that $R_{\star} \sin i$ is equal to $0.75 \pm 0.01$~R$_{\odot}$ and thus the inclination of TAP~4 equals $\sim$55$\degr$.

From Gaia's release, TAP~40's parallax value is 7.202 $\pm$ 0.016 mas\citep{gaia2021}, translating to a distance of 138.9 $\pm$ 0.3 pc or a distance modulus of 5.714 $\pm$ 0.005. The maximum V magnitude of TAP~40 is 12.5 \citep{grankin2008} and the unspotted V magnitude may be 12.3~$\pm$~0.2, assuming again a minimum spot coverage of ~20 per cent on the visible stellar surface. From the observed spectra we derived a $T_{\rm eff}$ of 4600~$\pm$~50~K and a $\log g$ of $4.6~\pm~0.2$. The observed averaged B-V is 1.09 $\pm$ 0.02 \citep{grankin2008} and the expected one for this effective temperature is 1.17 \citep{pecaut2013}, thus we can derive the visual extinction A$_{V}$ is equal to 0.25 $\pm$ 0.15. The bolometric correction is equal to -0.57 $\pm$ 0.05, and this results in an absolute bolometric magnitude of $5.77 \pm 0.25$. Thus, it can be derived that the luminosity of TAP~40 is $\log (L/L_{\odot}) = -0.41 \pm 0.1$. Then, we can infer that its stellar mass is $0.91 \pm 0.09$~M$_{\odot}$ and its radius is $0.9 \pm 0.1$~$R_{\odot}$. Compared with the evolutionary model of \citet{siess2000}, the age of TAP~40 is 28~Myr. The estimated $v \sin i$ of $12 \pm 0.2$ \kms\ implies that $R_{\star} \sin i$ is $0.35 \pm 0.01$ R$_{\odot}$ and thus the inclination of TAP~40 is about $\sim$25$\degr$. These adopted parameters of two stars are listed in Table \ref{tab:pars}.

\begin{table}
\caption{Parameters of TAP~4 and TAP~40.}
\label{tab:pars}
\begin{tabular}{lcc}
\hline
Parameters & TAP~4 & TAP~40\\
\hline
$M_{\star}$ (M$_{\odot}$)   & 0.95 $\pm$ 0.05 & 0.91 $\pm$ 0.09\\
$R_{\star}$ (R$_{\odot}$))  & 0.9  $\pm$ 0.1  & 0.9  $\pm$ 0.1 \\
Age (Myr)                   & $\sim$47 & $\sim$28\\
log g (cgs)                 & 4.6 $\pm$ 0.2 & 4.6 $\pm$ 0.2\\
$T_{\rm eff}$ (K)           & 5190 $\pm$ 50 & 4600 $\pm$ 50\\
$P_{\rm rot}$ (d)           & 0.482 & 1.585 \\
\vsini\ (\kms)              & 79 $\pm$ 0.5 & 12 $\pm$ 0.2 \\
inclination ($\degr$)       & $\sim$55 & $\sim$25 \\
Distance (pc)$^{a}$         & 145.4 $\pm$ 0.3 & 138.9 $\pm$ 0.3\\
\hline
$^a$ \citet{gaia2018,gaia2021}
\end{tabular}
\end{table}

\section{Zeeman-Doppler imaging}

We applied two imaging codes, \zdots\ \citep{hussain2000,hussain2001} and \zdi\ \citep{donati2006,donati2014}, to the time-series of Stokes I and V profiles to derive the surface brightness and magnetic field maps of TAP~4 and TAP~40. For the brightness map reconstruction, both codes allow cool and hot features to be recovered, as described by \citet{donati2014}. In the imaging process, we also derived the values of \vsini\ for the two stars and fine-tuned the rotational period for TAP~40, which are listed in Table \ref{tab:pars}.

Both imaging codes utilize the maximum entropy regularization \citep{skilling1984}, but the definitions of the entropy are slightly different in the two codes \citep{hussain2000}. Both codes implement a spherical harmonic decomposition of the magnetic field as described by \citet{donati2006} but with $\beta_{l,m}$ replaced by $\alpha_{l,m}+\beta_{l,m}$, as follows (see \citealt{lehmann2022,finociety2022}).

\begin{equation}
B_{r}(\theta,\phi) = -\sum_{l,m} \alpha_{l,m}Y_{l,m}(\theta,\phi)
\label{eq:br}
\end{equation}
\begin{equation}
B_{\theta}(\theta,\phi) = -\sum_{l,m} [(\alpha_{l,m}+\beta_{l,m})Z_{l,m}(\theta,\phi)+\gamma_{l,m}X_{l,m}(\theta,\phi)]
\end{equation}
\begin{equation}
B_{\phi}(\theta,\phi) = -\sum_{l,m} [(\alpha_{l,m}+\beta_{l,m})X_{l,m}(\theta,\phi)-\gamma_{l,m}Z_{l,m}(\theta,\phi)]
\end{equation}
where:
\begin{equation}
Y_{l,m} = c_{l,m}P_{l,m}(\cos\theta)e^{im\phi}
\end{equation}
\begin{equation}
Z_{l,m} = \frac{c_{l,m}}{l+1} \frac{\partial P_{l,m}(\cos\theta)}{\partial \theta}e^{im\phi}
\end{equation}
\begin{equation}
X_{l,m} = \frac{c_{l,m}}{l+1} \frac{P_{l,m}(\cos\theta)}{\sin\theta} im e^{im\phi}
\end{equation}
\begin{equation}
c_{l,m} = \sqrt{\frac{2l+1}{4\pi} \frac{(l-m)!}{(l+m)!}}
\end{equation}
where $P_{l,m}(\cos\theta)$ is the associated Legendre polynomial. Therefore, the imaging codes recover the complex coefficients of SH modes, $\alpha_{l,m}$, $\beta_{l,m}$ and $\gamma_{l,m}$, rather than the 3 components of the surface magnetic field for each independent pixel. This approach introduced a prior physical constraint on the reconstructed magnetic field map and allows one to easily derive the properties of poloidal and toroidal components of the surface magnetic field. Given the \vsini\ of the two stars, we limited $l_{\rm max}$ to be 15 and 7 for TAP~4 and TAP~40, respectively. We also performed imaging with larger $l_{\rm max}$, but only obtained marginal changes.

\subsection{Brightness and magnetic maps}

\begin{figure*}
\center
\includegraphics[width=.47\textwidth]{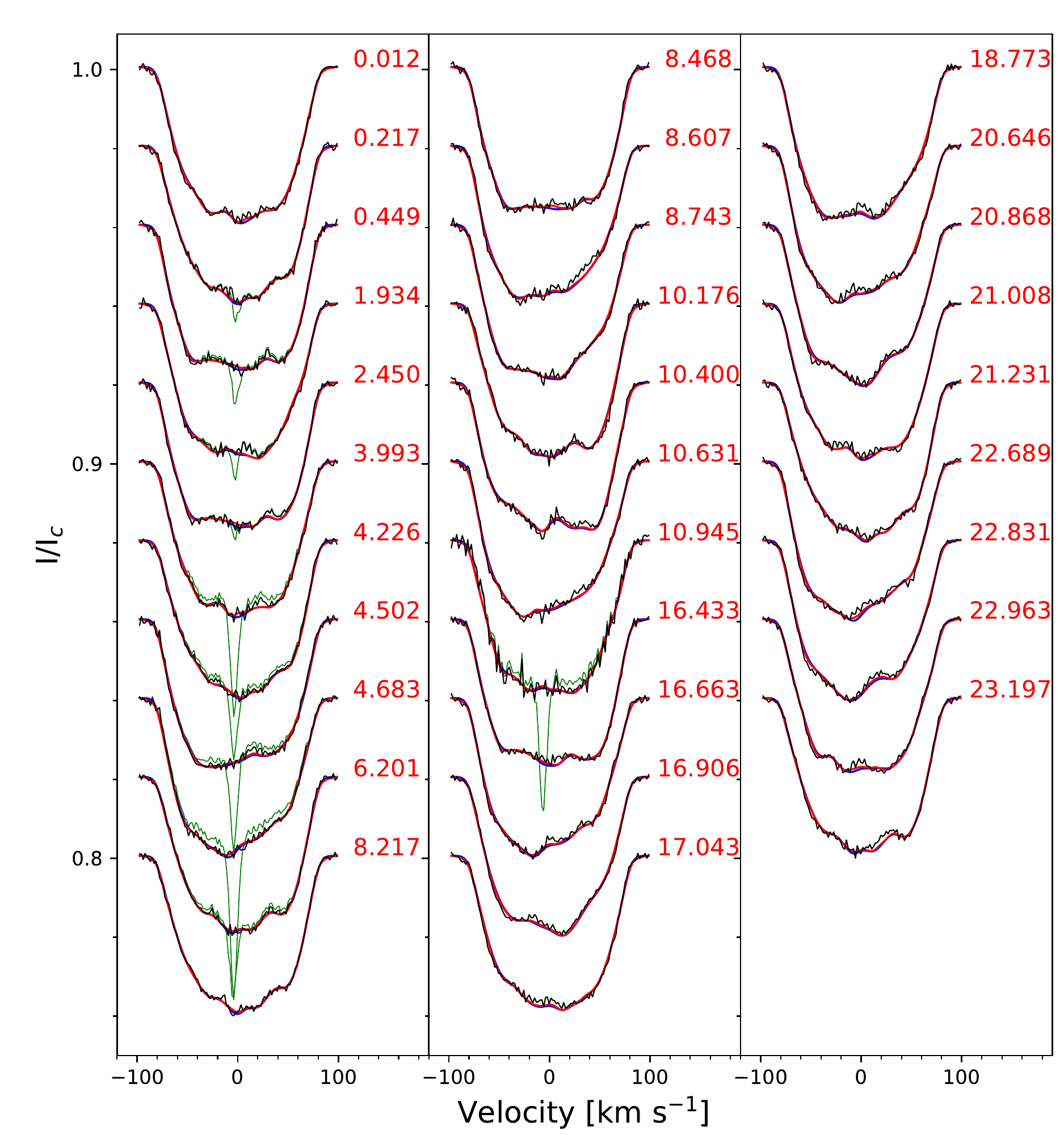}
\includegraphics[width=.47\textwidth]{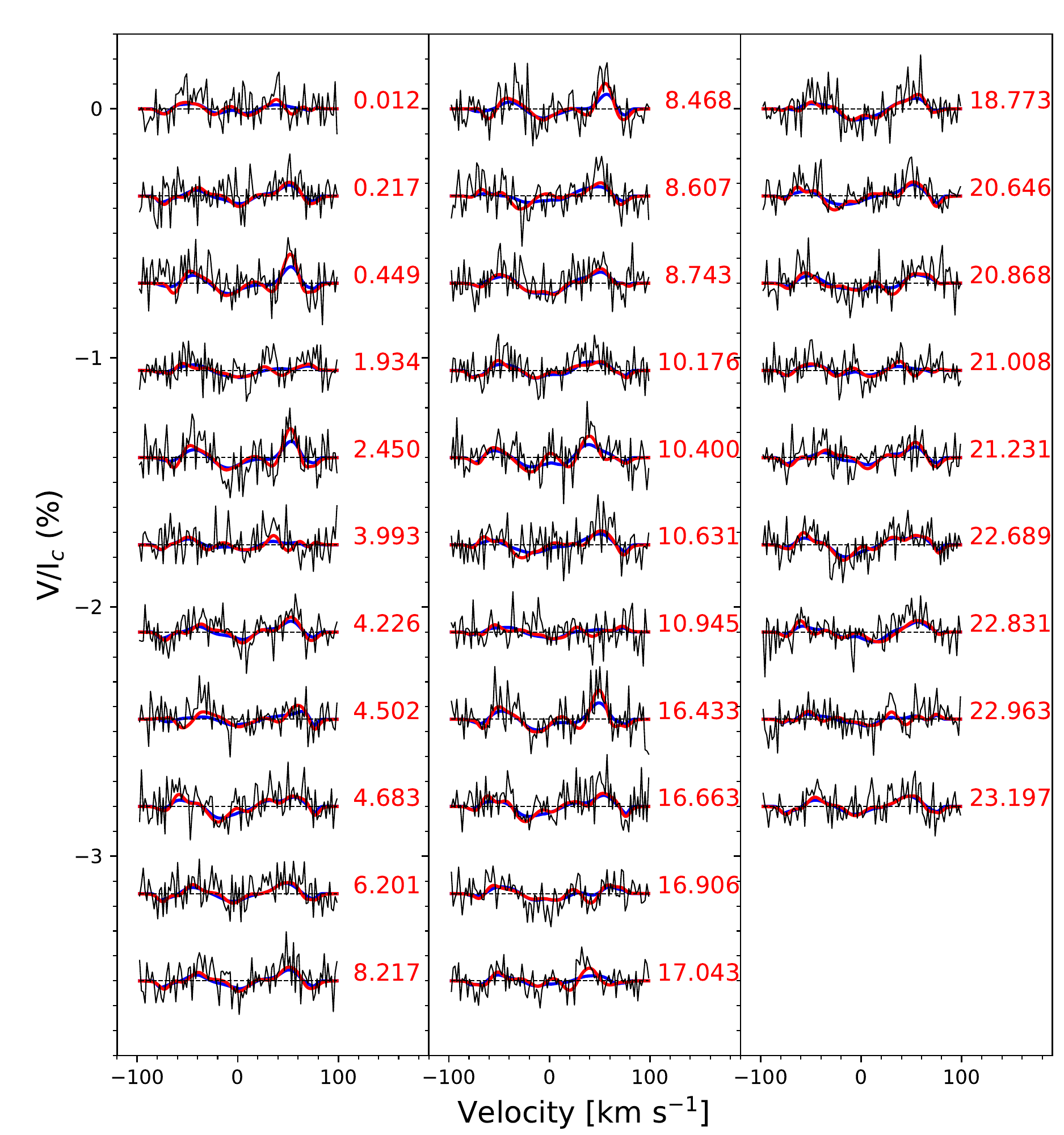}
\caption{Maximum entropy fits (blue lines for \zdots\ and red lines for \zdi) to the observed (black lines) Stokes I (left) and Stokes V (right) LSD profiles of TAP~4. The lunar contaminations are shown in green.}
\label{fig:sp_tap4}
\end{figure*}

\begin{figure}
\center
\includegraphics[width=.21\textwidth]{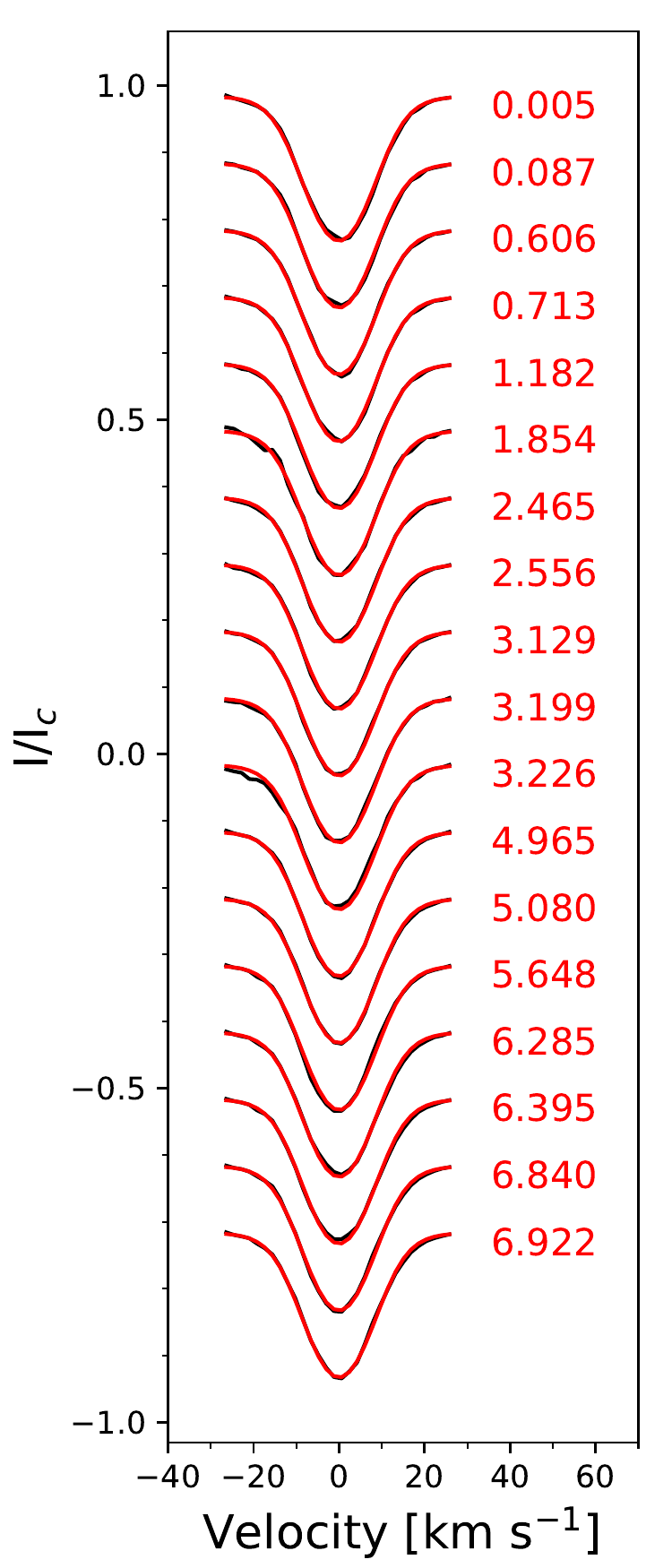}
\includegraphics[width=.21\textwidth]{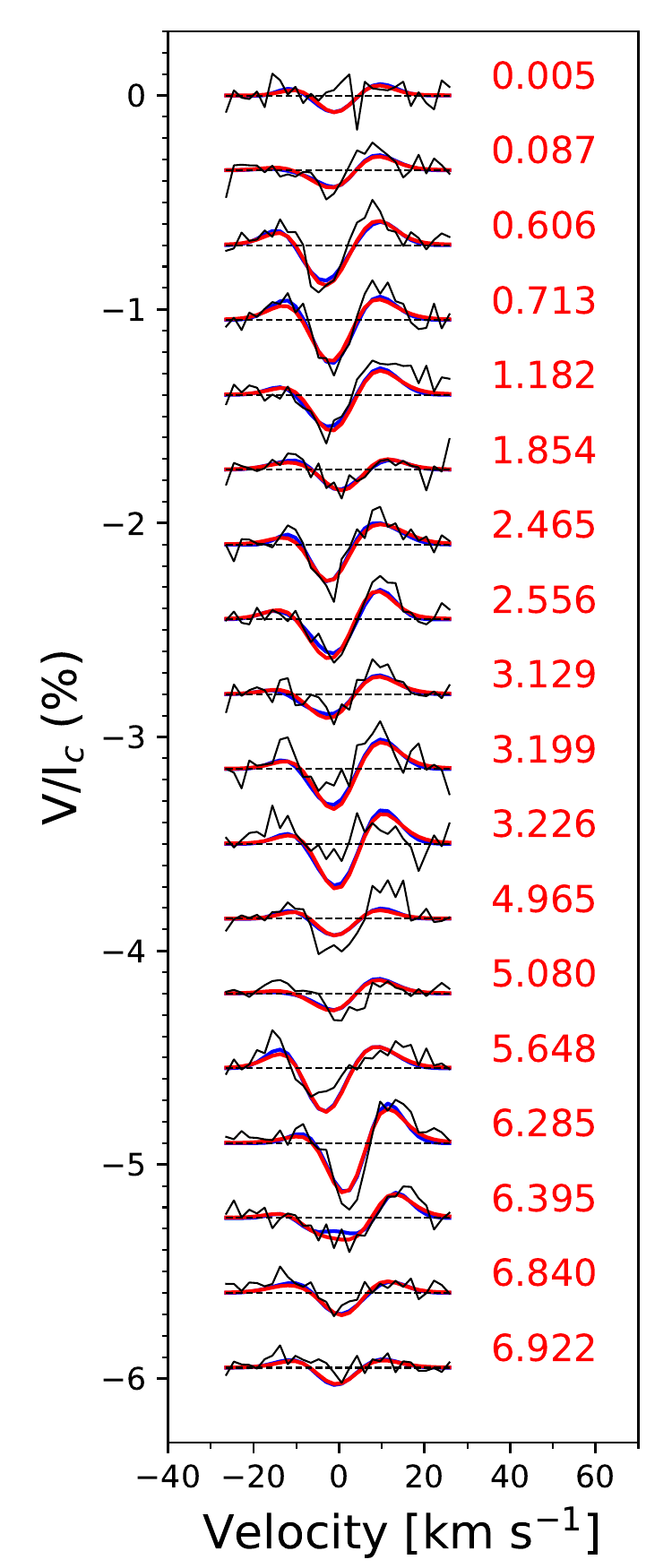}
\caption{Left panel shows the observed Stokes I profiles (black) and the immaculate profiles (red) for TAP~40. Right panel shows the maximum entropy fits derived by \zdots\ (blue) and \zdi\ (red) to the Stokes V profiles (black) of TAP~40.}
\label{fig:sp_tap40}
\end{figure}

We display the observed Stokes I and V profiles of TAP~4 as well as the maximum entropy fits derived by the two codes in Fig. \ref{fig:sp_tap4}, where the green lines in the Stokes I profiles represent the profiles before correction of the moon light contamination. For TAP 40, we find no clear modulated variations in its Stokes I profiles (see the left panel in Fig. \ref{fig:sp_tap40}), which means that surface brightness inhomogeneities of TAP~40, if any, have a low contrast with respect to the photosphere. The observed V profiles of TAP~40 and the corresponding fits to them are shown in the right panel of Fig. \ref{fig:sp_tap40}. As can be seen, the two codes give very consistent solutions to the observed profiles of TAP~4 and TAP~40. The fits correspond to the reduced chi-square $\chi_{r}^{2}$ of 1 for TAP~4 and 1.2 for TAP~40 by both codes. The brightness and magnetic maps of TAP 4 were derived separately from Stokes I and V profiles.

\begin{figure}
\center
\includegraphics[width=.4\textwidth]{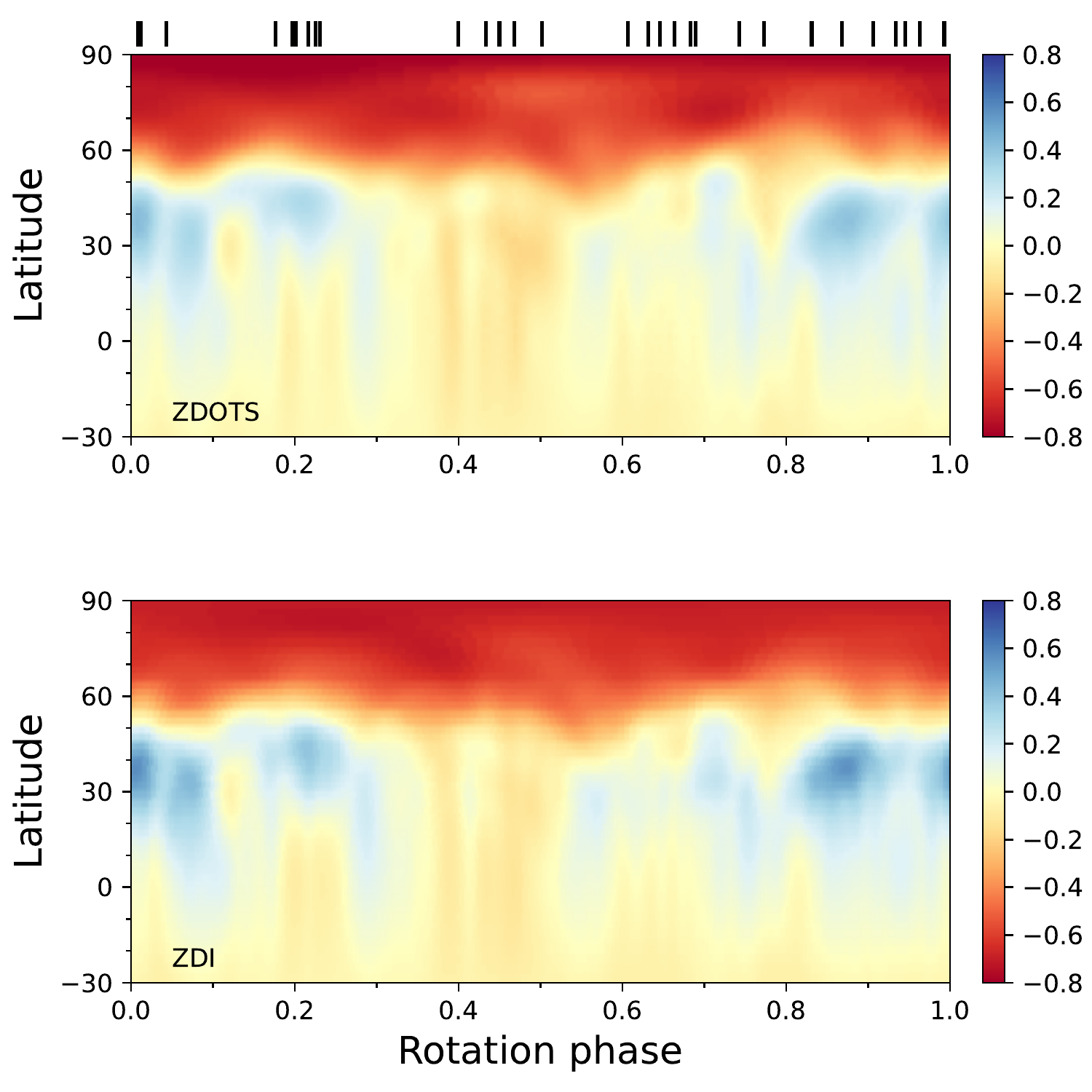}
\caption{Images of the logarithmic surface brightness relative to the photoshpere of TAP~4, produced by \zdots\ (top) and \zdi\ (bottom). For convenient comparison, both are rectangular and along the same longitudinal direction.}
\label{fig:bright}
\end{figure}

The brightness map reconstructions of TAP~4 derived by \zdots\ and \zdi\ are shown in Fig. \ref{fig:bright}. Both images show the logarithmic brightness relative to the photosphere. The surface features recovered by the two codes are nearly identical to each other.

The most prominent feature on the surface brightness map of TAP~4 is the large, cool polar spot with some appendages extending to low latitudes. This polar spot is related to the flat bottom of the observed Stokes I profiles (Fig. \ref{fig:sp_tap4}). In addition to the cool spots, TAP~4 also exhibits warm features at intermediate latitudes. The warm features shape a ring structure at around latitude 40$\degr$. The coverage of the cool spots on TAP~4 is 7 per cent and the value of the warm spots is 5 per cent. However, the coverage derived from the reconstructed image should be underestimated since Doppler imaging is only sensitive to the large-scale, inhomogeneous features. The small-sized spots and symmetric structures on the stellar surface can not be recovered by the Doppler imaging.

\begin{figure*}
\center
\includegraphics[width=.46\textwidth]{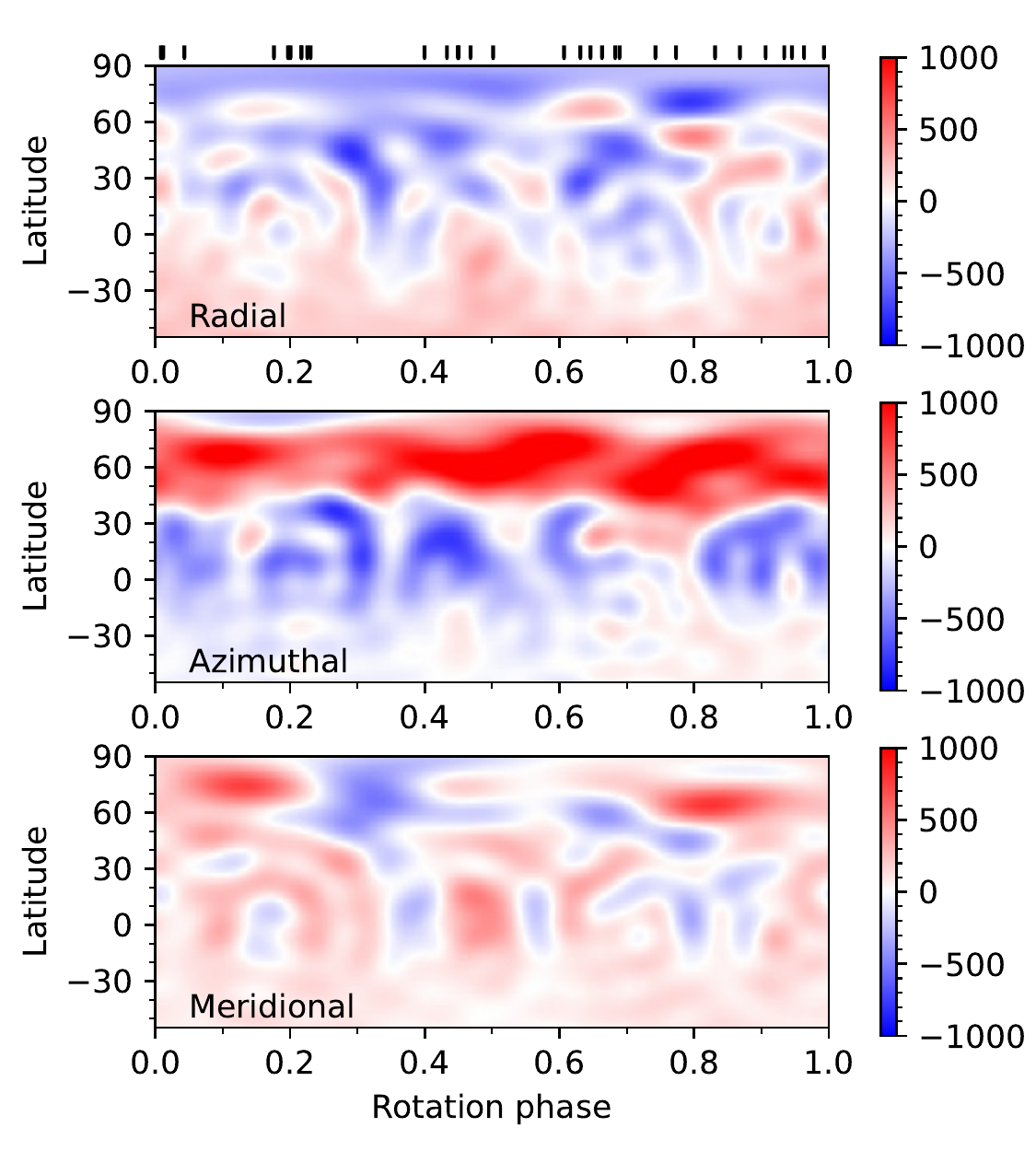}
\includegraphics[width=.46\textwidth]{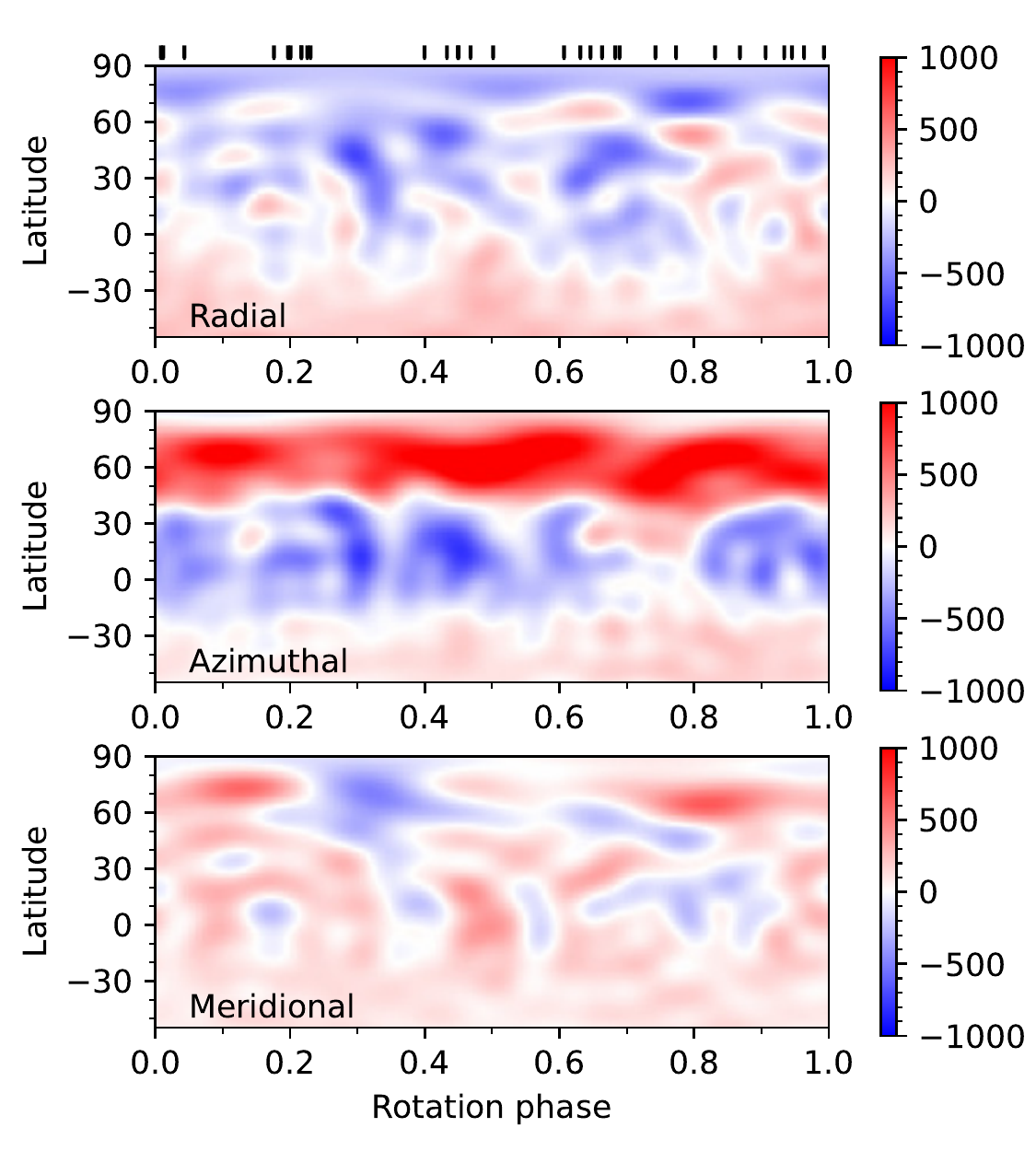}
\caption{Large-scale magnetic field maps of TAP~4 derived by \zdots\ (left) and \zdi\ (right).}
\label{fig:mag_tap4}
\end{figure*}

\begin{figure*}
\center
\includegraphics[width=.46\textwidth]{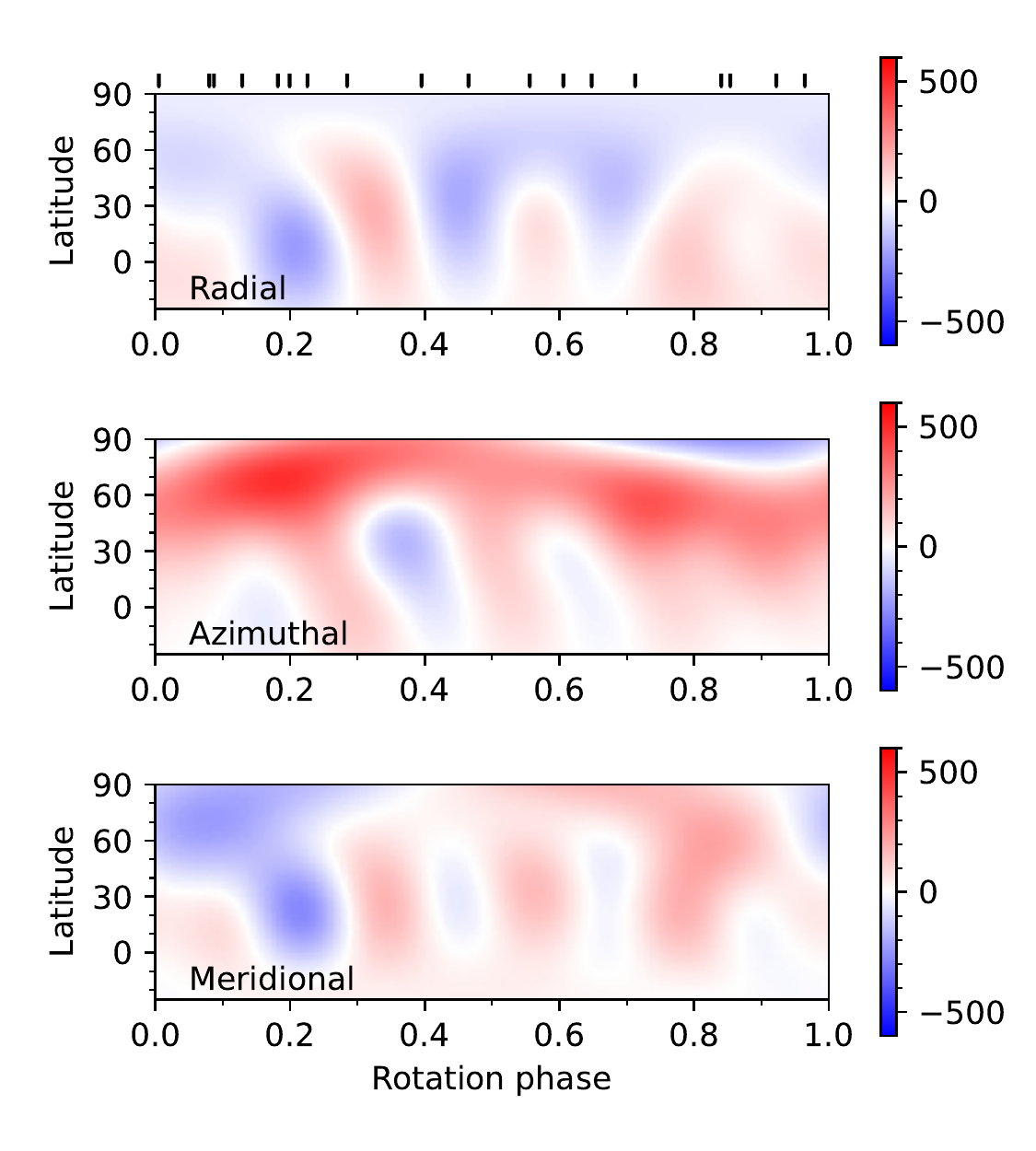}
\includegraphics[width=.46\textwidth]{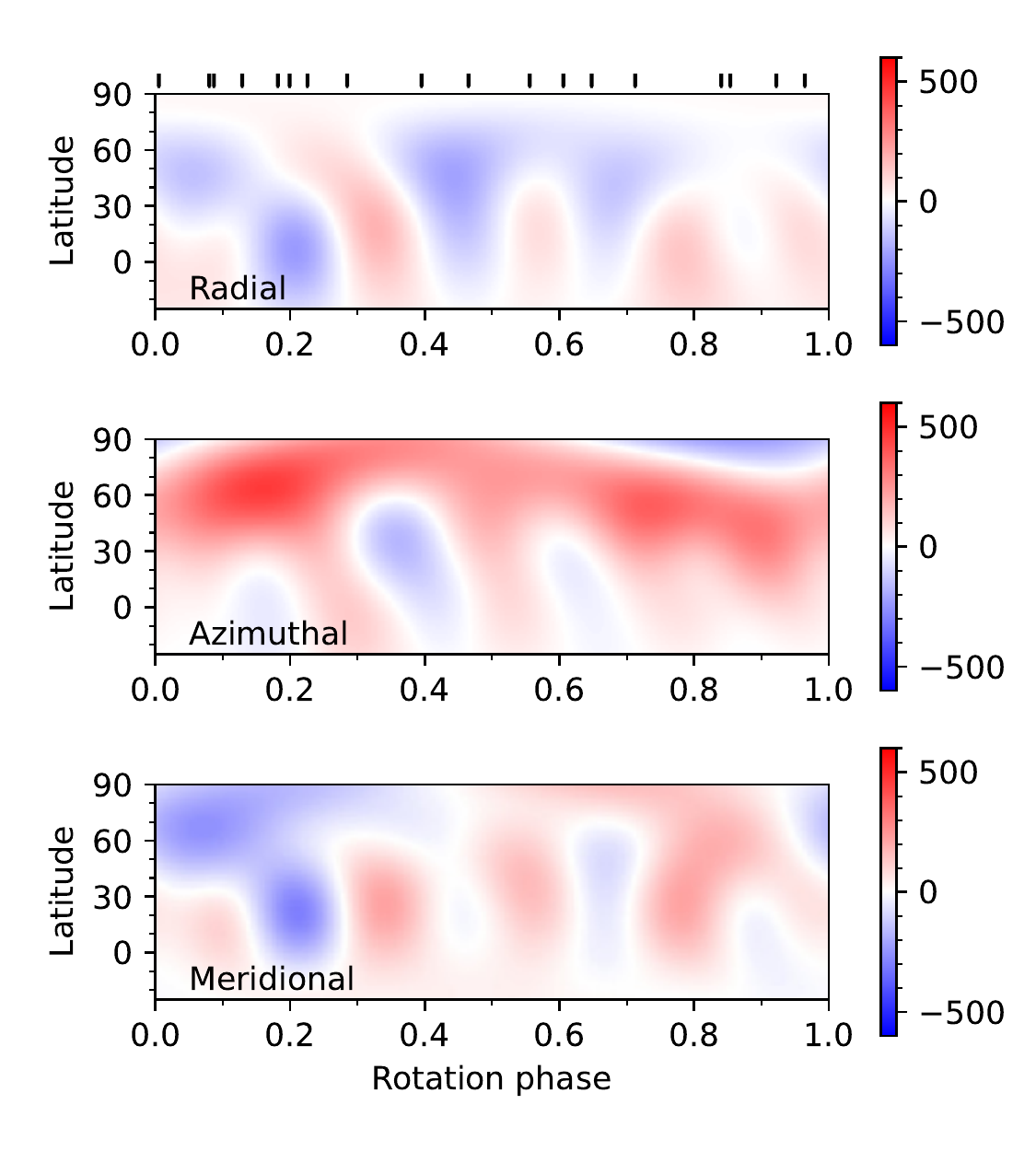}
\caption{Same as Fig. \ref{fig:mag_tap4}, but for TAP~40.}
\label{fig:mag_tap40}
\end{figure*}

The magnetic maps of TAP~4 and TAP~40 derived by \zdots\ and \zdi\ are also in good agreement, which are displayed in Fig. \ref{fig:mag_tap4} and \ref{fig:mag_tap40}, respectively. The properties of the magnetic field of TAP 4 and TAP 40 derived from the spherical harmonic coefficients obtained by the two codes are listed in Table \ref{tab:mag}. The reconstructed magnetic maps show a complex large-scale magnetic structures on the surface of TAP~4. The poloidal and toroidal components enclose 40 and 60 per cent of the total magnetic energy, respectively. The toroidal field of TAP~4 is dominated by the axisymmetric component, which is related to the strong positive ring structure of the azimuthal field as shown in the middle panel of Fig. \ref{fig:mag_tap4}. The radial field on TAP~4 is mainly dominated by the complex, non-axisymmetric, negative poloidal field. The large-scale dipole field of TAP~4 is tilted at $\sim$22\degr\ with respect to the rotation axis.

TAP~40 hosts a relatively simple large-scale magnetic field. The toroidal field is again the dominant component, enclosing more than 60 per cent of the total magnetic energy. This toroidal field shows up as a ring-like pattern in the azimuthal field map of TAP~40 in the middle panel of Fig. \ref{fig:mag_tap40}. The dipole field of TAP~40 is tilted at $\sim$42\degr\ with respect to the rotation axis. 

\begin{table}
\caption{Properties of the magnetic fields of TAP~4 and TAP~40 derived by the two codes.}
\label{tab:mag}
\begin{tabular}{lcc}
\hline
Parameters & \zdots\ & \zdi\\
\hline
TAP 4 &&\\
<B> (G)                             & 449  & 447 \\
poloidal (\% total)                 & 40   & 36  \\
dipole (\% poloidal)                & 37   & 42  \\
dipole tilt ($\degr$)               & 22   & 22  \\
dipole phase                        & 0.51 & 0.49\\
poloidal axisymmetric (\% poloidal) & 44   & 50  \\
toroidal (\% total)           & 60   & 64  \\
toroidal axisymmetric (\% toroidal) & 90   & 92  \\
&&\\
TAP 40 &&\\
<B> (G)                             & 168  & 163 \\
poloidal (\% total)                 & 34   & 36  \\
dipole (\% poloidal)                & 18   & 14  \\
dipole tilt ($\degr$)               & 43   & 42  \\
dipole phase                        & 0.34 & 0.33\\
poloidal axisymmetric (\% poloidal) & 15   & 14  \\
toroidal (\% total)           & 66   & 64  \\
toroidal axisymmetric (\% toroidal) & 71   & 66  \\
\hline
\end{tabular}
\end{table}

\begin{figure*}
\center
\includegraphics[width=.46\textwidth]{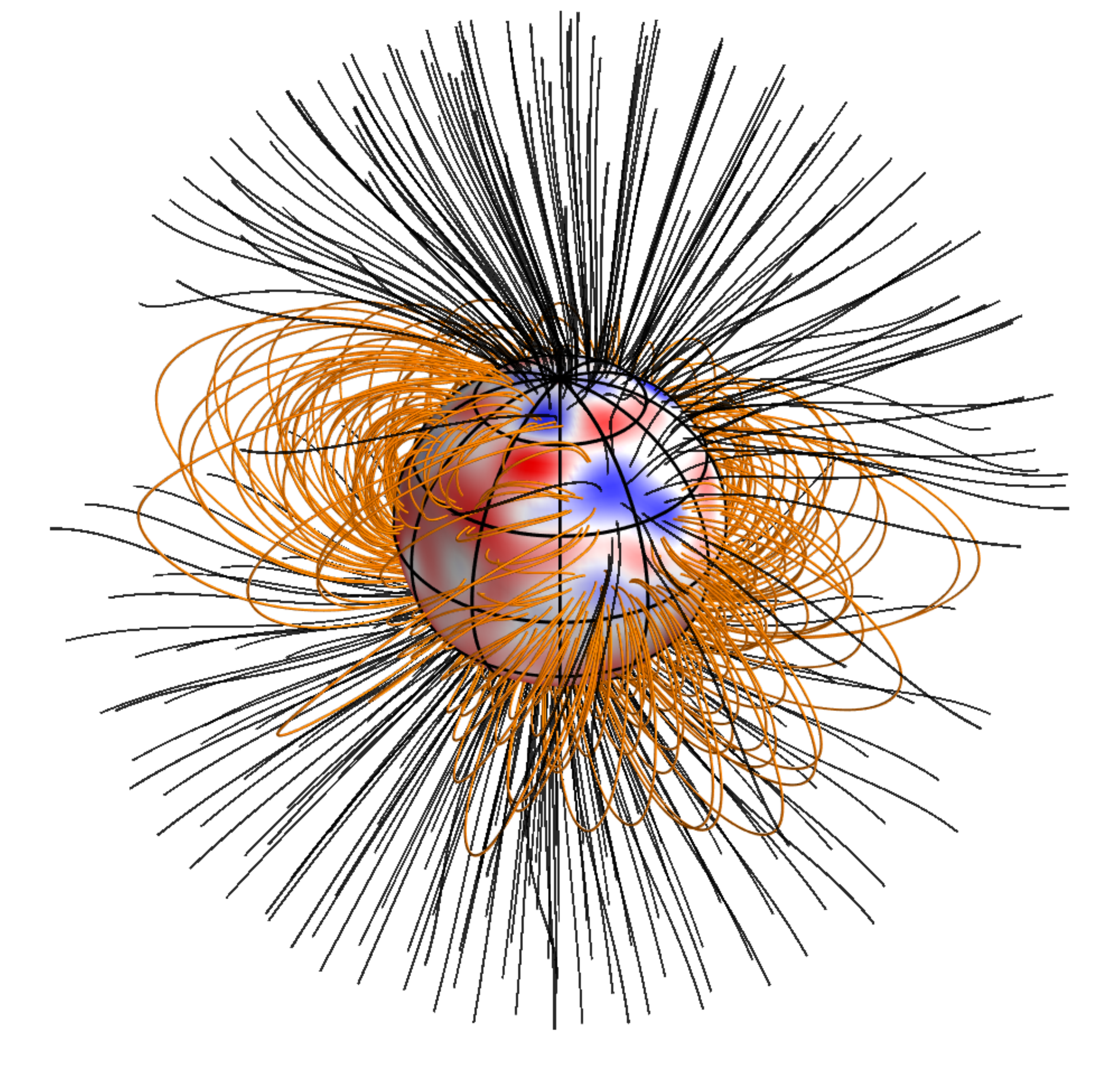}
\includegraphics[width=.46\textwidth]{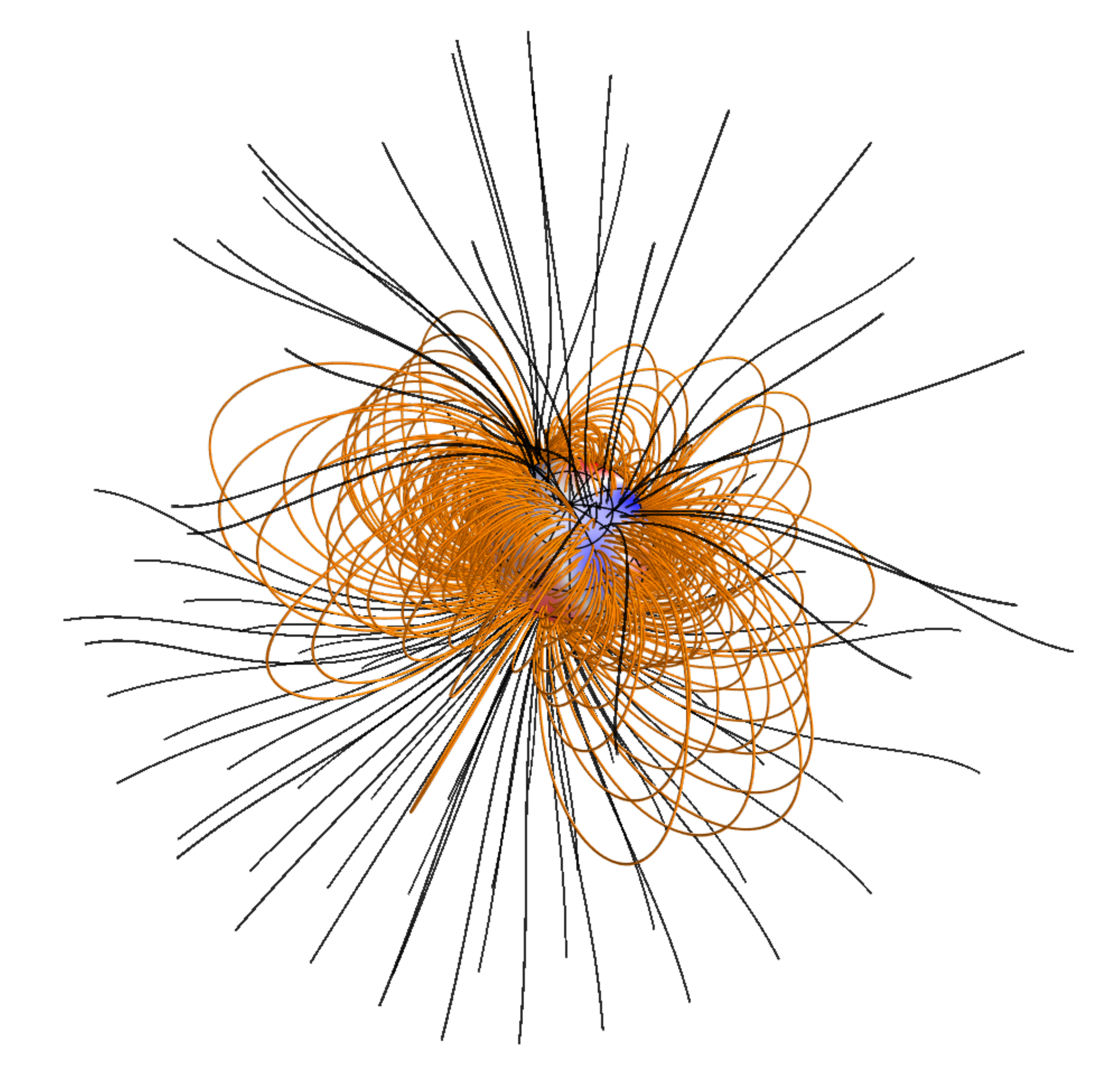}
\caption{Potential field extrapolations for TAP~4 and TAP~40, as viewed by an Earth-based observer, at phase 0.75. The orange lines represent the closed magnetic lines and the black lines denote the open magnetic lines. The source surfaces are set to 3~$R_{\star}$ for TAP~4 and 6~$R_{\star}$ for TAP 40, close to the corotation radii of the two stars.}
\label{fig:mag}
\end{figure*}

Based on the potential field source surface model presented by \citet{jardine2013}, we derived the extrapolation of the coronal magnetic field from the radial component of the reconstructed magnetic fields of TAP~4 and TAP~40. The source surfaces, where all magnetic field lines become radial, were set to 3~$R_{\star}$ for TAP~4 and 6~$R_{\star}$ for TAP~40, close to the corotation radii of these two stars respectively. The extrapolation results are shown in Fig. \ref{fig:mag}.

\subsection{Surface differential rotation}

Doppler imaging is also commonly used for the determination of surface differential rotation. The estimate can be achieved either by the cross-correlation of two reconstructed maps observed at close epochs or by looking for the differential rotation parameters that best fit the observed data \citep{donati1997b,donati2000,petit2002}. The shear method takes the surface differential rotation into account as part of the imaging process, which requires longer observations for differential rotation to be detectable with Doppler imaging.

The observations of TAP~4 span more than 23 rotation cycles, which makes the dataset well suited to estimate surface differential rotation. We thus applied the Doppler imaging code to the Stokes I profiles of TAP~4, with the assumption of a simple solar-like differential rotation law described by equation
\begin{equation}
\Omega(\theta) = \Omega_{\rm eq} - {\rm d}\Omega\sin^{2}\theta
\label{eq:dr}
\end{equation}
where $\Omega(\theta)$ is the rotation rate at latitude $\theta$ and $\Omega_{\rm eq}$ is the rotation rate at the stellar equator and ${\rm d}\Omega$ is the difference between rotation rates at the stellar equator and the pole. We derived Doppler images for various differential rotation parameter pairs of $\Omega_{\rm eq}$ and ${\rm d}\Omega$ at constant information content and recorded the resulting $\chi^{2}_{r}$. Then we fitted a paraboloid to the surface of $\chi^{2}_{r}$ distribution to find the parameter pair that leads to the minimum $\chi^{2}_{r}$ \citep{donati2003}.

\begin{figure}
\center
\includegraphics[width = .45\textwidth]{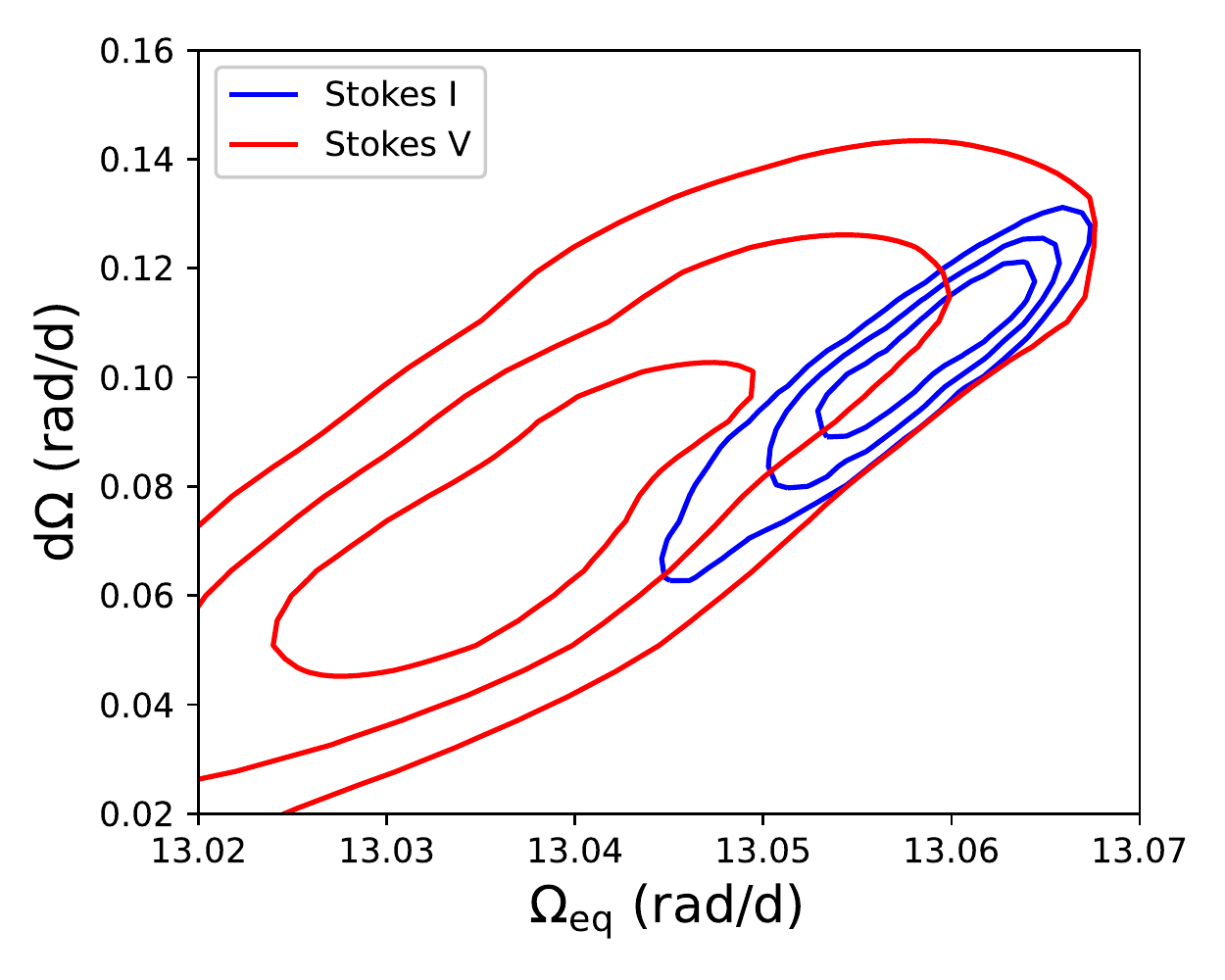}
\caption{Contour plot of $\chi^{2}_{r}$ versus $\Omega_{eq}$ and ${\rm d}\Omega$ pairs derived from the tomographic modelling on the Stokes I (blue) and V (red) profiles of TAP~4. The lines denote 1$\sigma$, 2$\sigma$ and 3$\sigma$ boundaries.}
\label{fig:diff}
\end{figure}

We show the contour plot of $\chi^{2}_{r}$ versus the differential rotation parameter pairs derived from the brightness and magnetic images of TAP 4 in Fig. \ref{fig:diff}. By fitting Stokes I profiles, we found a surface differential rotation rate of ${\rm d}\Omega = 0.105 \pm 0.007$ rad d$^{-1}$ and an equatorial rotation rate of $\Omega_{\rm eq} = 13.059 \pm 0.002$ rad d$^{-1}$ for TAP~4. From Stokes V profiles, we derived the surface shear rate of ${\rm d}\Omega = 0.065 \pm 0.014$ rad d$^{-1}$ and an equatorial rotation rate of $\Omega_{\rm eq} = 13.033 \pm 0.007$ rad d$^{-1}$.

\subsection{Filtering the activity jitter}

Young stars often exhibit high level of magnetic activity, which results in severe problems when trying to detect orbiting exoplanets through RV measurements. Magnetic activity can distort stellar spectral line profiles severely and thus induce large jitters in the RV curves of order a few of \kms\ \citep{donati2014,hebrard2014}. As a comparison, the RV amplitude of a low-mass star caused by the presence of a Jupiter-mass close-in close-in planet is of about 0.1~\kms\ (e.g. \citealt{donati2017}).

Another goal of the MaTYSSE programme is to detect close-in giant planets orbiting young active stars, which will provide important constraints for planet formation models. The technique proposed by \citet{donati2014} takes advantage of the tomographic modelling to filter the activity jitter in RV curves. Given the reconstructed brightness map of an active star, one can predict the RV jitter induced by surface activity and filter them out from the raw RV measurements so that the detection of small amplitude RV variation caused by a close-in giant planet becomes detectable. So far, hJs orbiting two young active stars, V830~Tau \citep{donati2016} and TAP~26 \citep{yu2017}, have been detected within the MaTYSSE programme.

\begin{figure*}
\center
\includegraphics[width = .85\textwidth]{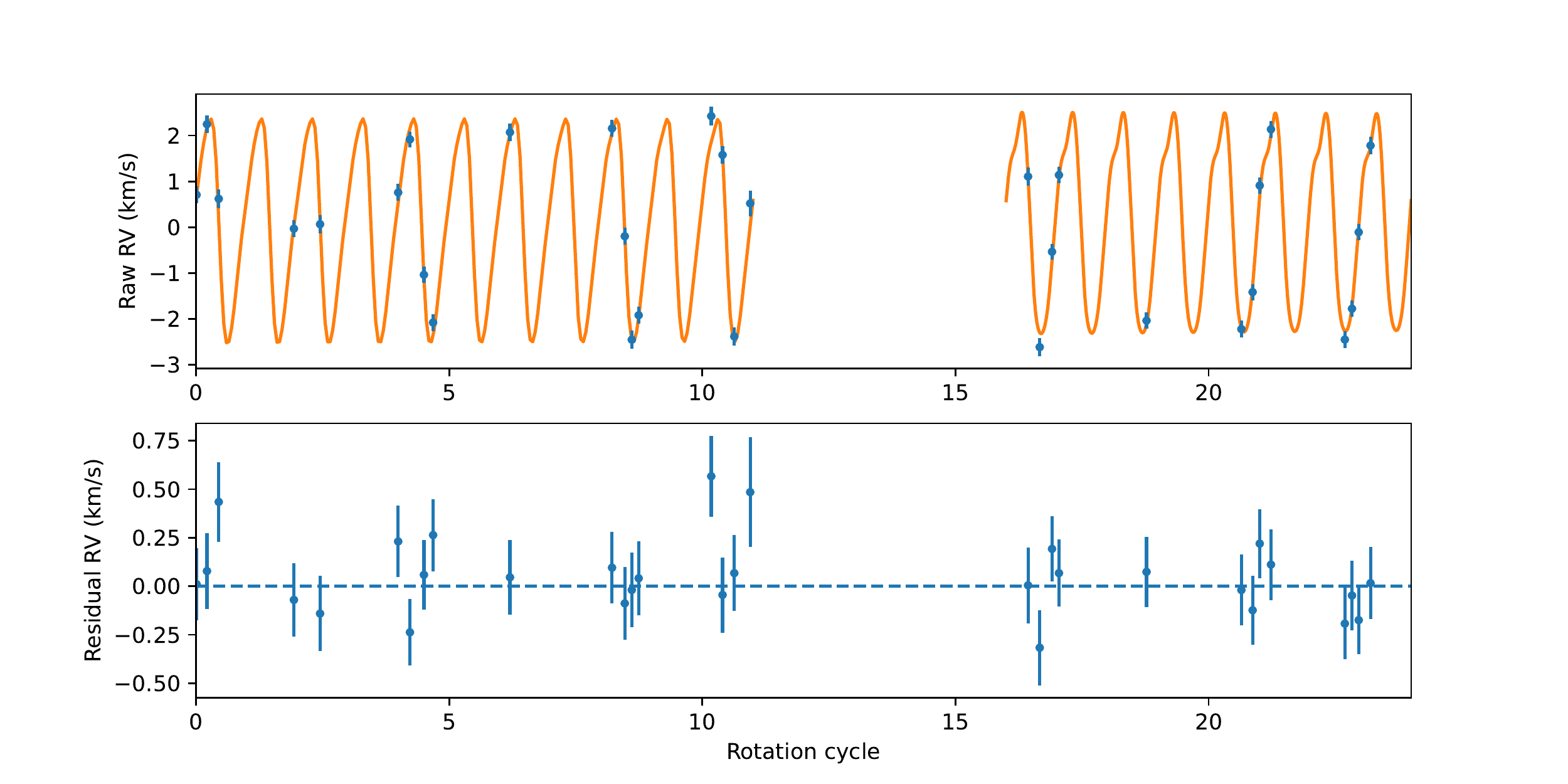}
\caption{Filtering the RV jitters induced by the magnetic activity for TAP~4.}
\label{fig:rv}
\end{figure*}

We performed the filtering process on the RV measurements of TAP~4, using the brightness map derived in Section~4.1. Due to the large gap within the observing run for TAP~4, we split the original dataset of TAP~4 into two subsets to reduce the rms caused by the evolution of the surface brightness map of TAP~4. The first subset spans rotation cycles 0--11 and the other one spans rotation cycles 16--24. The amplitude of the unfiltered RVs of TAP~4 is about 2 \kms, where the RVs of TAP~4 is derived from the first-order moment of the observed Stokes I profiles, rather than by a Gaussian fit due to the high \vsini\ and the distorted profile shapes. The RVs induced by the surface activity are shown in Fig. \ref{fig:rv} as orange lines. The rms dispersion of the residuals is 0.20 \kms, which is much larger than the intrinsic RV precision of ESPaDOnS of 0.03 \kms \citep{donati2014}. The high \vsini\ of TAP~4 strongly limits the accuracy of the filtering process. The mean 1$\sigma$ error of the RV measurements is 0.19~\kms for TAP~4 and the filtered RVs are not statistically different from the zero, which means TAP~4 is unlikely to host a close-in planet.

\section{Discussion and conclusions}

In this work, we presented a Zeeman-Doppler imaging study of two young active stars, TAP~4 and TAP~40, based on new high-resolution spectropolarimetric data collected with ESPaDOnS at  CFHT in November 2013, within the MaTYSSE large programme.

The analysis on the new high-resolution spectroscopic data shows that TAP~4 has a $T_{\rm eff}$ of 5190~$\pm$~50~K and a $\log g$ of 4.6~$\pm$~0.2 and TAP~40 has a $T_{\rm eff}$ of 4600 K~$\pm$~50~K and a $\log g$ of 4.6~$\pm$~0.2. TAP~4 is a rapidly rotating star with a \vsini\ of 79~$\pm$~0.5~\kms\ whereas TAP~40 is a relatively slower rotator with a \vsini\ of 12~$\pm$~0.2~\kms. Based on the parallax values from the Gaia astrometric solution \citep{gaia2018} and the photometric data of \citet{grankin2008}, we obtain the locations of these two stars on the HR diagram (Fig. \ref{fig:evl}). Combining with the evolutionary tracks of \citet{siess2000}, we find that TAP~4 has a stellar mass of 0.95~$\pm$~0.05$M_{\odot}$ and a radius of 0.9~$\pm$~0.1$R_{\odot}$ and TAP~40 has a mass of 0.91~$\pm$~0.09$M_{\odot}$ and a radius of 0.9~$\pm$~0.1$R_{\odot}$. The ages of TAP~4 and TAP~40 are 47~Myr and 28~Myr, respectively. Therefore, TAP~4 is probably at the late stage of wTTS phase and very close to the zero-age main sequence (ZAMS).

We applied two Zeeman-Doppler imaging codes to the time-series of Stokes I and V profiles to reconstruct their surface brightness and magnetic field images. However, the time-series of Stokes I profiles of TAP~40 show minimal variations, meaning that brightness features have a low contrast with respect to the photosphere at the surface of this star. The reconstructed surface brightness and magnetic field maps of both stars derived by the two different codes are in very good agreement with each other.

The brightness map of TAP~4 shows a pronounced, cool polar spot with some dark appendages extending to lower latitudes. The cool spot is not symmetric with respect to the rotation pole. In addition, the presence of warm spots is found at latitude 40$\degr$ on TAP~4. Considering the high \vsini\ and the short rotation period of TAP~4, the polar spot is expected by the current magnetic flux-tube models (e.g. \citealt{isik2007}). Similar polar features are commonly detected on the rapidly rotating stars in the sample of the MaTYSSE programme, e.g. TWA6 \citep{hill2019}, V530~Persei \citep{cang2020} and V471~Tau \citep{zaire2021}. Meanwhile, similar intermediate-latitude warm features are found on wTTSs with different parameters, e.g. LkCa~4 \citep{donati2014}, V830~Tau \citep{donati2016} and TAP~26 \citep{yu2019}. We may infer that the configuration of the surface brightness map of TAP~4 is very common among wTTSs.

TAP~40 shows very weak variations in its Stokes I profiles, which is consistent with the photometric results of \citet{grankin2008}. In their analysis, TAP~40 only exhibited a mean V band photometric amplitude of 0.109 mag, though their photometric observations were at different epochs. The small variations in the unpolarized profiles of TAP~40 are likely due to the low inclination.

The large-scale magnetic field of TAP~4 is more complex than that on TAP~40, which may be partly due to the lower \vsini\ of TAP~40 resulting in the lower resolution of its Zeeman-Doppler image. However, the most prominent features of the magnetic fields of these two stars are similar, which is a strong ring pattern of the azimuthal field seen on both surfaces of TAP~4 and TAP~40, as shown in Fig. \ref{fig:mag_tap4} and Fig. \ref{fig:mag_tap40}. Both of the magnetic reconstructions of TAP~4 and TAP~40 indicate predominant toroidal fields enclosing about 60 per cent of the total magnetic field energy. The dominance of the toroidal field is consistent with the inference of \citet{petit2008}, who showed that the rapid rotators with short periods are more likely to host dominant toroidal components. The ratio of the poloidal and toroidal fields is dependent on the Rossby number \citep{donati2009}.

Thanks to the long duration of the observing run for TAP~4, which covers more than 23 rotation cycles, it allows us to derive the surface differential rotation rate of TAP~4. Under the simple assumption of a solar-like differential rotation law, we obtain a surface shear of $0.105 \pm 0.007$~rad~d$^{-1}$ from Stokes I profiles and  $0.065 \pm 0.014$~rad~d$^{-1}$ from Stokes V profiles. In literature, the solar-like surface differential rotation is detected on various PMS to ZAMS stars, e.g. V471~Tau \citep{hussain2006,zaire2021,zaire2022} and HD 155555 \citep{dunstone2008}. The rapidly rotating ZAMS star AB~Dor even shows variations in the surface differential rotation with years \citep{donati2003,jeffers2007}. The value of the surface shear we derived for TAP~4 is relatively common among the wTTSs sample of the MaTYSSE programme. Differential rotation is one of the key ingredients of stellar dynamos, which amplify and sustain the stellar magnetic fields. Estimates of surface differential rotation on young active stars bring constrains on stellar dynamo models for stars at early evolutionary stages.

The tomographic modelling of TAP~4 is also used for filtering the activity-induced RV jitter to search for the potential close-in massive exoplanet. The rms of the filtered RVs we achieve is 0.2 \kms\ from an initial dispersion of raw RVs of 1.7 \kms. The dispersion is higher than that for other wTTSs in the MaTYSSE programme, e.g. LkCa~4 \citep{donati2014} but similar to that for TWA~6 \citep{hill2019}. The larger uncertainty is due to the relatively high \vsini\ of TAP~4. We do not find obvious deviations of the filtered RVs of TAP~4, and the filtered RV curve does not show any significant periodic signal. RV simulations can be used to derive the detection limit on the mass of potential exoplanets \citep{yu2019,finociety2021,nicholson2021}. We simulated the RV signatures from planets with various mass at a distance of 0.1~au assuming a circular orbit and found that an exoplanet more massive than $\sim$3.5~M$_{\rm Jup}$ orbiting TAP~4 would have been detectable at a 3$\sigma$ level with our data. With a false alarm probability smaller than 1 per cent, we can thus conclude that no close-in planet more massive than 3.5~M$_{\rm Jup}$ orbits TAP 4 at a distance of about 0.1~au.

\section*{Acknowledgements}
This paper is based on observations obtained at the Canada-France-Hawaii Telescope (CFHT), operated by the National Research Council of Canada, the Institut National des Sciences de I'Univers of the Centre National de la Recherche Scientifique (INSU/CNRS) of France and the University of Hawaii. This study is supported by the National Natural Science Foundation of China under grants Nos.10373023, 10773027, U1531121, 11603068 and 11903074. We acknowledge the science research grant from the China Manned Space Project with NO. CMS-CSST-2021-B07. JFD acknowledges funding from the European Research Council (ERC) under the H2020 research \& innovation programme (grant agreement \#740651 NewWorlds).

\section*{Data availability}
The raw data used in this work from the CFHT can be obtained at the CADC (https://www.cadc-ccda.hia-iha.nrc-cnrc.gc.ca/en/cfht/)

\bibliographystyle{mnras}
\bibliography{tap4andtap40}

\label{lastpage}
\end{document}